\documentclass[journal,comsoc]{IEEEtran}
\usepackage{graphicx}
\usepackage{subfigure}
\usepackage{float}
\usepackage{subfloat}
\usepackage{amsmath}
\usepackage{amssymb}
\usepackage{amsthm}
\usepackage{epstopdf}
\usepackage{cases}
\usepackage{color}
\usepackage{xcolor}
\usepackage{makecell}
\usepackage{cite}
\usepackage{booktabs}
\usepackage{diagbox}
\usepackage{multirow}
\usepackage{stfloats}
\usepackage{algorithmic}
\usepackage[ruled]{algorithm2e}
\usepackage{cases}

\usepackage[normalem]{ulem}

\usepackage{caption}
\usepackage{mathtools}
\usepackage[T1]{fontenc}
\ifCLASSINFOpdf

\else

\fi
\usepackage{amsmath}
\begin{document}

\title{Time-Varying Offset Estimation for Clock-Asynchronous Bistatic ISAC Systems}

\author{Yi~Wang,~\IEEEmembership{Member,~IEEE,}
        Keke~Zu,~\IEEEmembership{Member,~IEEE,}
        Luping~Xiang,~\IEEEmembership{Senior~Member,~IEEE,}
        Martin~Haardt,~\IEEEmembership{Fellow,~IEEE}
        and~Kun~Yang,~\IEEEmembership{Fellow,~IEEE}
\thanks{This work was supported in part by the Municipal Government of Quzhou under Grant (No. 2023D027, 2023D045, 2023D005), and in part by the Natural Science Foundation of China under Grant No. 62132004. (\emph{Corresponding author: Yi Wang.})}
\thanks{Y. Wang and K. Zu are with the Yangtze Delta Region Institute (Quzhou), University of Electronic Science and Technology of China, Quzhou 324003, China, e-mail: wangyi@csj.uestc.edu.cn, zukeke@csj.uestc.edu.cn.}
\thanks{L. Xiang is with the State Key Laboratory of Novel Software Technology, Nanjing University, Nanjing 210008, China, and also with the School of Intelligent Software, Engineering, Nanjing University (Suzhou Campus), Suzhou 215163, China, e-mail: luping.xiang@nju.edu.cn.}
\thanks{M. Haardt is with the Communications Research Laboratory, Ilmenau University of Technology, Ilmenau, Germany, email: martin.haardt@tu-ilmenau.de.}
\thanks{K. Yang is with the State Key Laboratory of Novel Software Technology, Nanjing University, Nanjing, 210008, China, and School of Intelligent Software and Engineering, Nanjing University (Suzhou Campus), Suzhou, 215163, China, and School of Information and Communication Engineering, University of Electronic Science and Technology of China, Chengdu 611731, China, e-mail: kunyang@essex.ac.uk.}
}

\maketitle

\begin{abstract}
The bistatic Integrated Sensing and Communication (ISAC) is poised to become a key application for next generation communication networks (e.g., B5G/6G), providing simultaneous sensing and communication services with minimal changes to existing network infrastructure and hardware. However, a significant challenge in bistatic cooperative sensing is clock asynchronism, arising from the use of different clocks at far separated transmitters and receivers. This asynchrony leads to Timing Offsets (TOs) and Carrier Frequency Offsets (CFOs), potentially causing sensing ambiguity. Traditional synchronization methods typically rely on static reference links or GNSS-based timing sources, both of which are often unreliable or unavailable in UAV-based bistatic ISAC scenarios. To overcome these limitations, we propose a Time-Varying Offset Estimation (TVOE) framework tailored for clock-asynchronous bistatic ISAC systems, which leverages the geometrically predictable characteristics of the Line-of-Sight (LoS) path to enable robust, infrastructure-free synchronization. The framework treats the LoS delay and the Doppler shift as dynamic observations and models their evolution as a hidden stochastic process. A state-space formulation is developed to jointly estimate TO and CFO via an Extended Kalman Filter (EKF), enabling real-time tracking of clock offsets across successive frames. Furthermore, the estimated offsets are subsequently applied to correct the timing misalignment of all Non-Line-of-Sight (NLoS) components, thereby enhancing the high-resolution target sensing performance. Extensive simulation results demonstrate that the proposed TVOE method improves the estimation accuracy by 60\%.
\end{abstract}


\IEEEpeerreviewmaketitle
\section{Introduction}
\IEEEPARstart{T}{he} Integrated Sensing and Communication (ISAC) technology presents an effective solution to achieve load saving and spectrum reuse by using unified transceiver and spectrum resources, while providing efficient cooperative detection and sensing data sharing capabilities to support functions such as positioning, ranging, velocimetry, imaging, detection, and identification, which can effectively alleviate the problem of insufficient detection capability of single radar sensors, and realize the improvement of the overall performance of the network and business capabilities \cite{int1,int2,int3,int4}. For example, ISAC can sense and track low-altitude Unmanned Aerial Vehicles (UAVs), and avoid UAVs from intruding into specific areas by establishing electronic fences, which is challenging to achieve in a cost-efficient and easy-to-implement manner \cite{int5,int6,int7}.

Conventional sensing mainly refers to two modes: monostatic sensing and bistatic sensing. Monostatic sensing refers to the configuration where the transmitter and receiver are co-located \cite{int8,int9}. Implementing sensing functions into communication systems using full-duplex radio is an innovative approach that leverages the dual capabilities of communication hardware, but it needs sophisticated interference cancellation technology to avoid self-interference. Besides, in current mobile networks operating in Time-Division Duplexing (TDD) mode, an extra receiver needs to be added with physical separation from the transmitter on the same Base Station (BS) \cite{int10}. This requirement substantially increases the size, cost, and complexity of the BS. Additionally, the self-interference from the BS's own transmission is much stronger than the reflected signal, necessitating suppression through digital cancellation techniques \cite{int11,int12}. As a result, deploying monostatic sensing in mobile networks poses a significant challenge for operators. Bistatic sensing is more promising for the practical implementation of ISAC because it requires almost no change to the existing network infrastructure \cite{int14,int15}. The transmitter and receiver are largely separated spatially, alleviating the significant demands imposed by full-duplex technology. However, their clocks, driven by local oscillators, are typically not locked or synchronised, potentially causing clock deviation. With the clock asynchronism, the Timing Offsets (TOs) and Carrier Frequency Offsets (CFOs) are imposed, thereby leading to time-varying phase offsets in the measurements of Channel State Information (CSI) \cite{int16,int17}. This issue hinders the coherent processing of discontinuous CSI measurements and can introduce ambiguities in the estimation of delay and Doppler frequency.

Although conventional OFDM systems can estimate and mitigate TO and CFO using known pilots or preambles at the receiver side, these approaches rely on the assumption that pilot sequences are perfectly known and received over a reliable, single-path channel. Unlike data communications, where offsets are generally manageable, offset mitigation in sensing applications poses a significant challenge. For instance, a clock stability of 20 parts per million can result in a maximum timing offset of 20 nanoseconds over 1 millisecond, resulting in a ranging error of up to 6 meters. Recently, new techniques have emerged to tackle the issue of clock asynchronism. The authors of reference \cite{CACC1} introduced the Cross-Antenna Cross-Correlation (CACC) method for passive sensing by leveraging the conjugate cross-correlation between pairs of antennas. The authors of references \cite{CACC2,CACC3} employed the CACC method to resolve sensing ambiguities, enabling a static UE and BS to form an uplink bistatic system for environmental sensing. Similar to CACC, Cross-Antenna Signal Ratio (CASR) utilizes the ratio between two antennas to mitigate the asynchronism factors. The authors of reference \cite{CASR1} proposed to employ the absolute CASR over symbols for Doppler frequency estimation. Another approach in \cite{CASR2} involved estimating the Doppler frequency based on the properties of the CSI ratio, such as its periodicity and the differences or correlations between segments of CSI ratio signals. Additionally, a Taylor series-based method was introduced to utilize the CSI ratio for multi-target sensing, deriving the second-order complex Taylor series of the CSI ratio in a closed-form expression relative to the vector of sensing parameters to be estimated \cite{CASR3}. However, the sensing accuracy of these methods relies on the quality of the constructed reference signal, which may inevitably introduce errors and degrade the performance of bistatic sensing.


Accurate synchronization is essential for bistatic ISAC systems, particularly in enabling precise range and velocity estimation. In the recent literature, many synchronization techniques have focused on leveraging the Line-of-Sight (LoS) path or other static propagation components to estimate timing and frequency offsets. For example, \cite{int15,Sta1,Sta2,Sta3} utilize the delay of the strongest path--assumed to be the LoS--as a temporal reference, applying either cross-correlation with known preambles or delay-profile matching to estimate the TO. Similarly, carrier frequency offsets are typically estimated by exploiting the phase evolution across subcarriers or OFDM symbols, assuming quasi-static channels. While such methods have shown success in fixed infrastructure or low-mobility scenarios, their applicability to UAV-based systems is limited. In these settings, the LoS path may experience fast time-varying delays and Doppler shifts due to UAV motion, or may even be intermittently blocked. As a result, synchronization methods based on static channel assumptions may fail to maintain accuracy and robustness under UAV mobility and oscillator drift.

To overcome this challenge, we propose a dynamic synchronization framework that treats the LoS path as a continuously observable reference signal. By tracking its geometric delay and Doppler across frames, we form a virtual synchronization sequence and apply sequential filtering to estimate and predict clock offsets. This proposed approach, termed Time-Varying Offset Estimation (TVOE), is inherently adaptive to mobility, resilient to partial LoS degradation, and capable of compensating clock-induced misalignments across all Non-Line-of-Sight (NLoS) components. The main contributions of this paper are summarized below:

\begin{itemize}
\item We propose a TVOE framework tailored for bistatic ISAC systems involving mobile UAV nodes, where synchronization is challenged by mobility-induced time-varying offsets and the absence of a shared clock. By leveraging the geometric predictability of the LoS path, the TVOE achieves robust offset estimation even in dynamic scenarios where traditional methods become unreliable or inapplicable.
\item A time-varying state-space model is formulated to jointly estimate the TO and the CFO, treating the delay and Doppler of the LoS path as noisy, time-dependent observations. An Extended Kalman Filter (EKF) is then employed to enable real-time, drift-aware tracking of the clock offsets under UAV-induced dynamics and oscillator instability.
\item  By applying offset correction at the NLoS level, the proposed framework restores temporal coherence across multipath returns, thereby enabling high-precision target localization and motion estimation even in the presence of severe clock asynchronism and UAV-induced channel dynamics.
\end{itemize}

\begin{figure}[!t]
	\centering
	\includegraphics[width=0.35\textheight]{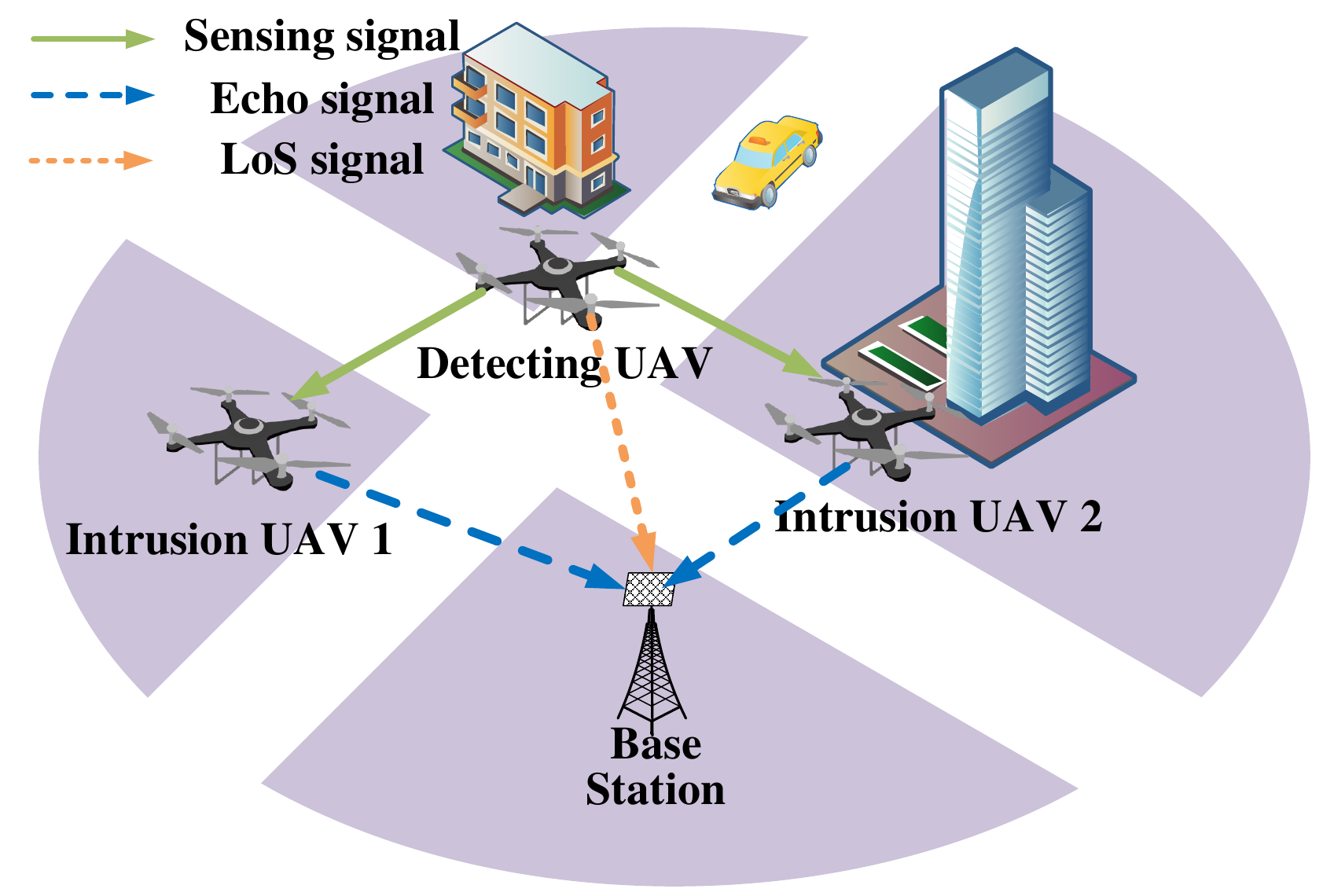}
	\DeclareGraphicsExtensions.
	\caption{The bistatic ISAC scenario.}
	\label{scen}
\end{figure}

The remainder of this article is structured as follows: Section II presents the system model of the bistatic ISAC system. Section III outlines the proposed TVOE framework. Section IV details the performance analysis of the estimators. The networking performance of the bistatic ISAC system is examined in Section V, with results provided. Finally, Section VI concludes the article.

\textbf{Notation}: Bold uppercase letters denote matrices (e.g., $\boldsymbol{A}$). Bold lowercase letters denote column vectors (e.g., $\boldsymbol{v}$). Scalars are denoted by normal font (e.g., $x$); For an arbitrary matrix $\boldsymbol{A}$, rank($\boldsymbol{A}$), tr($\boldsymbol{A}$), $\boldsymbol{A^T}$, $\boldsymbol{A^H}$, and $\boldsymbol{A_{p,q}}$ denote its rank, trace, transpose, conjugate, and the $p$-th row and $q$-th column element, respectively. Moreover, ${\left\| {\cdot} \right\|}$ and ${\left| {\cdot} \right|}$ denote the Euclidean distance and the magnitude, respectively. The operators $ \otimes $, $ \odot $, and $ \circ $ denote the Kronecker product, Khatri-Rao product, and the Hadamard product, respectively.

\section{System Model}

\subsection{Scenario}
Consider a bistatic cooperative sensing scenario in which a ground BS and a detecting UAV conduct bistatic sensing operations, as shown in Fig.~\ref{scen}. The detecting UAV acts as an aerial base station, transmitting ISAC signals that include pilot and private data information. These signals are reflected by nearby intrusion UAVs and collected by the ground BS for simultaneous data transmission and target detection. The signals for communication are conveyed directly and do not contain any target sensing information, while the sensing signals reflected by surrounding intrusion UAVs are employed to estimate the targets' distance, speed, and angle through passive bistatic sensing.

We assume that the BS and the detecting UAV are physically static and that a dominating Line-of-Sight (LoS) path exists between them; specifically, the distance between them and the AoA of the LoS path at the receiver are known, and the path parameters remain approximately stable during a short Coherent Processing Interval (CPI). Additionally, both the BS and the UAV are assumed to operate in half-duplex mode. Assume that the BS and the detecting UAV are equipped with two Uniform Planar Arrays (UPAs) for transmitting and receiving with half-wavelength antenna spacing, respectively. The size of UPA is $M_x \times M_y$. The three-dimensional position vectors of the BS and detecting UAV is represented as $\boldsymbol{{q_b}} = {\left[ {{x_b},{y_b},{z_b}} \right]^T}$ and $\boldsymbol{{q_u}\left( t \right)} = {\left[ {{x_u}\left( t \right),{y_u}\left( t \right),{z_u}\left( t \right)} \right]^T}$, respectively. The UAV moves along a trajectory $\boldsymbol{{q_u}}\left( \Delta t \right) = \boldsymbol{{q_0}} + \boldsymbol{v} \Delta t + \boldsymbol{\Delta q}\left( \Delta t \right)$, where $\boldsymbol{{q_0}}$ denotes the initial position of the UAV, $\boldsymbol{v} \in \mathbb{R}^{3}$ denotes its average velocity vector in 3D Cartesian space over the time interval $\Delta t$, and $\boldsymbol{\Delta q}\left( \Delta t \right)$ represents small deviations from ideal linear motion due to dynamic disturbances.

\subsection{Signal Model}
After propagation through the multipath channel, the ISAC signals transmitted from the BS contain both LoS signals, conveyed directly for communication, and NLoS signals, which are reflected by surrounding targets for radar sensing simultaneously. By leveraging the pilot signal for both communication channel estimation and passive bistatic sensing, the system achieves a multi-functional use of the transmitted signals.

As the Orthogonal Frequency Division Multiplexing (OFDM) signal is widely used in current 5G systems and will likely also be used in 6G systems, we employ OFDM as the transmitted signal model of the proposed ISAC system. The general baseband transmit signal is expressed as \cite{SM1}
\begin{equation}
s\left( t \right) = \sum\limits_{m = 0}^{M - 1} {\sum\limits_{n = 0}^{N - 1} {{s_{m,n}} \cdot {e^{j2\pi \left( {{f_c} + n\Delta f} \right)t}}\text{rect}\left( {\frac{{t - mT_s}}{T_s}} \right)} }
\end{equation}
where $M$ and $N$ denote the number of OFDM symbols and the number of subcarriers, respectively; ${s_{m,n}}$ denotes the transmit baseband symbol on the $n$-th subcarrier of the $m$-th OFDM symbol, ${f_c}$ denotes the carrier frequency, ${\Delta f}$ denotes the subcarrier spacing and $\text{rect}\left( x \right)$ denotes the rectangle function. The subcarrier interval is ${\Delta f} $ and the OFDM symbol period is ${T_s} = \frac{1}{{\Delta f}} + {T_g}$ where ${T_g}$ is the period of cyclic prefix.

The transmit signal can be represented as \cite{SM2}
\begin{equation}
\boldsymbol{x \left( t \right)} = \boldsymbol{{w_t}}s\left( t \right),
\end{equation}
where $\boldsymbol{{w_t}}$ denotes the transmit beamforming vector. The beamforming vectors for communication and sensing can be generated by using a generalized Least Square (LS) method respectively according to the desired beam patterns. Then a multi-beam design for ISAC can be generated by combining the two beamforming vectors through a phase shifting term ${e^{j\varphi }}$ and a power distribution factor $0 \le {\beta _R} \le 1$. Hence, the beamforming vectors $\boldsymbol{{w_t}}$ can be written as
\begin{equation}
\boldsymbol{{w_t}} = \sqrt {{\beta _R}} {e^{j\varphi }}{\boldsymbol{w_{t,s}}} + \sqrt {1 - {\beta _R}} {\boldsymbol{w_{t,c}}},
\end{equation}
where $\boldsymbol{{w_{t,s}}}$, $\left\| {\boldsymbol{{w_{t,s}}}} \right\|_2^2 = 1$, and $\boldsymbol{{w_{t,c}}}$, $\left\| {\boldsymbol{{w_{t,c}}}} \right\|_2^2 = 1$ denote two beamforming vectors of the scanning and fixed sub-beams, respectively.

\begin{figure}[!t]
	\centering
	\includegraphics[width=0.2\textheight]{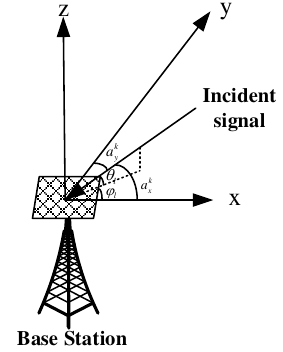}
	\DeclareGraphicsExtensions.
	\caption{The azimuth angle and elevation angle of AoA.}
	\label{angle}
\end{figure}

Since UAVs typically operate at relatively high altitudes, the communication links between the detecting UAV and the BS are generally dominated by the LoS component. Therefore, the assumption that a dominant LoS path always exists between them is reasonable. According to the Rician channel model for UAV-enabled ISAC networks \cite{SM3}, it can be written as
\begin{equation}
\boldsymbol{{H}} = \frac{{\sqrt {{K_R}} }}{{\sqrt {{K_R} + 1} }}\boldsymbol{H^{LoS}} + \frac{1}{{\sqrt {{K_R} + 1} }}\boldsymbol{H^{NLoS}},
\end{equation}
where
\begin{equation}
{\boldsymbol{H^{LoS}} = {\beta _{0}}{e^{j2\pi {f_{0}}t}}\delta \left( {t - {\tau _{0}}} \right)\boldsymbol{\alpha \left( {{q}_{r,0}} \right){\alpha ^T}\left( {{q}_{t,0}} \right)}}
\end{equation}
and
\begin{equation}
{\boldsymbol{H^{NLoS}} = \sum\limits_{l = 1}^L {{\beta _{l}}{e^{j2\pi {f_{l}}t}}\delta \left( {t - {\tau _{l}}} \right)\boldsymbol{\alpha \left( {{q}_{r,l}} \right){\alpha ^T}\left( {{q}_{t,l}} \right)} }}
\end{equation}
denote the LoS component and NLoS components of the channel, respectively, and $\boldsymbol{{H}} \in {\mathbb{C}^{{M_x^u  M_y^u }  \times  {M_x^b  M_y^b } }}$. The constant ${K_R}$ denotes the Rician K-factor, ${L}$ denotes the number of multipaths, and ${{\beta _{l}}}$, ${{f_{l}}}$, ${{\tau _{l}}}$ denote the channel fading coefficient, the Doppler frequency, and the delay of the $l \in \left\{ {0,1,2, \cdots ,{L}} \right\}$-th path, respectively; $\boldsymbol{{{q}_{r,l}}}$ and $\boldsymbol{{{q}_{t,l}}}$ are the Angle of Arrival (AoA) and Angle of Departure (AoD), respectively. Moreover, ${\beta _{0}} = \sqrt {\frac{{{\lambda ^2}}}{{{{\left( {4\pi {d_{L}}} \right)}^2}}}}$ and ${\beta _{l}} = \sqrt {\frac{{{\lambda ^2}}}{{{{\left( {4\pi } \right)}^3}d_{NL,1}^2d_{NL,2}^2}}} {\rho _{l}}$ denote the attenuations of the LoS path and NLoS paths, respectively; ${\rho _{l}}$ denotes the reflecting factor of $l$-th path, accounting for the signal attenuation due to surface reflectivity. Here, $d_{NL,1}$ and $d_{NL,2}$ represent the distances from the transmitter to the reflector and from the reflector to the receiver, respectively. For an array of $M_x^b \times M_y^b$ antennas with an AoA/AoD $\boldsymbol{{{{q}}_k}} = {\left( {{\varphi _k},{\theta _k}} \right)^T}$, the array steering vector can be expressed as
\begin{equation}
\boldsymbol{\alpha\left( {{{{q}}_k}} \right)} = \left[ {\begin{array}{*{20}{c}}
1\\
{{e^{-j\pi \Omega _y^k}}}\\
 \vdots \\
{{e^{-j\left( {M_y^b - 1} \right)\pi \Omega _y^k}}}
\end{array}} \right] \otimes \left[ {\begin{array}{*{20}{c}}
1\\
{{e^{-j\pi \Omega _x^k}}}\\
 \vdots \\
{{e^{-j\left( {M_x^b - 1} \right)\pi \Omega _x^k}}}
\end{array}} \right],
\end{equation}
where $\Omega _y^k = \cos a_{y}^k =  \sin {\theta _k}\sin {\varphi _k}$, $\Omega _x^k = \cos a_{x}^k =  \sin {\theta _k}\cos {\varphi _k}$ are the combination of sine and cosine functions of direction, the operator $\otimes$ denotes the Kronecker product, $a_{x}^k$ and $a_{y}^k$ are the angle between the incident signal and the x and y axes, respectively, and ${\varphi _k}$ and ${\theta _k}$ are the azimuth angle and elevation angle of the AoA as shown in Fig.~\ref{angle}, respectively.

The received signal for either sensing or communication is thus given by
\begin{equation}
\begin{array}{l}
\begin{aligned}
&\boldsymbol{{y_{n,m}}} = \boldsymbol{{H_{n,m}}{w_t}}{s_{n,m}} + \boldsymbol{{z_{n,m}}}\\
&= \sum\limits_{l = 0}^L {{\beta _l}{\chi _{t,l}}{e^{ - j2\pi n\Delta f{\tau _l}}} \cdot {e^{j2\pi {f_l}m{T_s}}} \cdot } \boldsymbol{\alpha \left( {{q_{r,l}}} \right)}{s_{n,m}} + \boldsymbol{{z_{n,m}}}
\end{aligned}
\end{array},
\end{equation}
where $\boldsymbol{{H_{n,m}}}$ denotes the channel response on the $n$-th subcarrier of the $m$-th packet, ${\chi _{t,l}} = \boldsymbol{{\alpha ^T}\left( {{q_{t,l}}} \right){w_t}}$ denotes the the transmit beamforming gain, and $\boldsymbol{{z_{n,m}}}$ denotes the Additive White Gaussion Noise (AWGN).


As for the sensing, after propagation through the multipath channel, the motion state of targets can be estimated using bi-static sensing. Since the signal strength after multiple reflections is negligible, we only consider echoes that are directly reflected from the targets. For simplicity, we assume that only one channel estimation is performed on each packet and $M_s$ CSI estimates are obtained for bistatic sensing at an interval of $T_p^s$ where $M_s$ denotes the number of OFDM packets. We also assume that there are ${P_s}$ OFDM symbols in each packet and the interval of CSI estimation is $T_p^s = {P_s}{T_s}$. Notably, the directly conveyed communication path does not convey any sensing information.

\section{Time-Varying Offset Estimation for Bistatic ISAC Systems}
Owing to the spatial separation between the transmitter and receiver, their clocks--driven by independent local oscillators--are typically not synchronized. This lack of synchronization leads to clock deviation, which manifests itself as TOs and CFOs, severely impairing the performance of sensing and communication functions in bistatic ISAC systems.
To address the challenges of clock asynchronism in UAV-enabled bistatic ISAC systems, we propose a Time-Varying Offset Estimation (TVOE) algorithm, which dynamically estimates and tracks TOs and CFOs using only the observable LoS component.
The core idea of TVOE is to treat the LoS path as a natural synchronization reference whose delay and Doppler shift evolve smoothly over time. By comparing the measured LoS parameters (e.g., delay and Doppler) with their predicted geometric counterparts, a residual sequence is constructed that reflects the underlying clock-induced deviations. These residuals are modeled as noisy observations of a hidden stochastic process, and sequential filtering techniques--such as the EKF--are employed to recursively track the TO and CFO across frames. Unlike conventional synchronization methods that rely on static reference paths or predefined pilot structures, the TVOE explicitly adapts to the UAV mobility and clock drift, enabling robust and continuous offset tracking even in the absence of external anchors or shared pilot signals.

The TVOE framework is a lightweight, adaptive, and receiver-side synchronization mechanism designed specifically for UAV-enabled bistatic ISAC systems. It aims to jointly estimate and track TOs and CFOs induced by oscillator drift and UAV mobility. Unlike conventional approaches that require GNSS-based synchronization or dedicated pilot exchange between transmitter and receiver, the TVOE operates in a fully passive and infrastructure-free manner, relying only on the received ISAC signals and predictable properties of the LoS path. As illustrated in Fig. 2, the TVOE framework comprises the following three tightly coupled components:

a) LoS-based residual construction: The receiver compares the measured LoS delay and the Doppler against geometry-derived predictions based on the UAV trajectory. The resulting residuals serve as clean, high-SNR observations of the clock-induced offset process.

b) Sequential offset estimation via EKF: A dynamic state-space model is formulated where the TO and the CFO evolve over time. These offsets are recursively tracked via an EKF, enabling frame-level synchronization that adapts to time-varying conditions.

c) Offset compensation for NLoS paths: The TO and CFO estimates obtained from the LoS path are applied to the entire received signal, effectively aligning all multipath components. This global compensation enables high-resolution sensing by restoring delay-Doppler coherence and mitigating distortion that would otherwise impair weak NLoS returns.

By leveraging only the predictable behavior of the LoS path, the TVOE framework eliminates the dependency on external timing infrastructure and significantly improves synchronization robustness under dynamic, infrastructure-less, or adversarial conditions, rendering it a strong candidate for future 6G UAV-based ISAC deployments.

\subsection{LoS-Based Residual Construction}
In UAV-enabled bistatic ISAC systems, the TVOE framework exploits the geometrically predictable nature of the LoS path between the UAV and BS. Unlike NLoS components that are affected by complex scattering and reflection mechanisms, the LoS path is typically strong, temporally persistent, and geometrically predictable--especially when the UAV's trajectory is known or can be accurately measured.
As the UAV moves along a predictable or measurable flight path, the time-varying propagation characteristics of the LoS component--specifically its delay and Doppler shift--can be accurately predicted based on the system's geometry. This predictability forms the foundation for constructing residual sequences that reflect the underlying clock asynchronism, enabling reliable offset tracking in dynamic and infrastructure-less environments.


Given the known UAV trajectory, the expected delay of the LoS path can be approximated as
\begin{equation}
\tau _{\rm LoS}^{\rm geo} \left( t \right) = \frac{{\left\| {\boldsymbol{{q_u}}\left( t \right) - \boldsymbol{{q_b}}} \right\|}}{c},
\end{equation}
where $\boldsymbol{{q_b}}$ denotes the location of the base station.

Similarly, the Doppler shift of the LoS path at time $t$ can be estimated based on the relative velocity between the UAV and the base station along the line-of-sight direction as
\begin{equation}
f_{\rm LoS}^{\rm geo}\left( t \right) \approx \frac{1}{\lambda } \cdot \frac{d}{{dt}}\left\| {\boldsymbol{{q_u}}\left( t \right) - \boldsymbol{{q_b}}} \right\|.
\end{equation}

Let ${{\hat{\tau}} _{\rm{LoS}}\left( t \right)}$ and ${{\hat{f}} _{\rm{LoS}}\left( t \right)}$ denote the measured delay and Doppler shift of the LoS path at frame $t$, extracted via range-Doppler processing techniques. Then, the residuals between the measured and predicted values are defined as ${\tau _{\rm{off}}\left( t \right)} = {{\hat{\tau}} _{\rm{LoS}}\left( t \right)} - \tau _{\rm LoS}^{\rm geo} \left( t \right)$ and ${f _{\rm{off}}\left( t \right)} = {{\hat{f}} _{\rm{LoS}}\left( t \right)} - f_{\rm LoS}^{\rm geo}\left( t \right)$.


These residuals capture the discrepancies introduced by clock asynchronism--specifically, the TOs and CFOs. Since the LoS path is strong and minimally affected by multipath fading, the residuals offer a stable and high-SNR observation of the underlying clock drift process. By collecting ${\tau _{\rm{off}}\left( t \right)}$ and ${f _{\rm{off}}\left( t \right)}$ over successive time slots, we obtain a temporal sequence of LoS-based residuals. This sequence serves as a virtual synchronization reference, enabling continuous and anchor-free tracking of the TO and the CFO using a state-space estimation model, as elaborated in the next section.


\subsection{Offset Estimation via EKF}
The core of the TVOE framework lies in recursively estimating the latent timing and frequency offsets (TO and CFO) from the LoS-based residual sequence. Since these offsets vary smoothly over time due to local oscillator drift and UAV motion, we formulate a dynamic state-space model to describe their temporal evolution and apply sequential Bayesian filtering (i.e., EKF) to track them across frames.

We define the state vector at time $t$ as
\begin{equation}
\boldsymbol{{x_t}} = \left[ {\begin{array}{*{20}{c}}
{{\tau _{\rm off}}\left( t \right)}\\
{{f_{\rm off}}\left( t \right)}
\end{array}} \right]
\end{equation}

We assume the state evolves over time as a random walk model with additive Gaussian drift
\begin{equation}
\boldsymbol{{x_{t + 1}}} =  \boldsymbol{{x_t}} + \boldsymbol{{u_t}},
\end{equation}
where $\boldsymbol{{u_t}}  \sim \mathcal{N}\left( {0, \boldsymbol{Q_x}} \right)$ is the process noise with covariance $\boldsymbol{Q_x}$, representing the gradual drift due to unsynchronized clocks and UAV dynamics.

The measurement at each time $t$ is the residual offset obtained from LoS-based residual construction
\begin{equation}
\boldsymbol{{z_t}} = \left[ {\begin{array}{*{20}{c}}
{{\hat{\tau} _{\rm off}}\left( t \right)}\\
{{\hat{f}_{\rm off}}\left( t \right)}
\end{array}} \right] = \boldsymbol{{x_t}} + \boldsymbol{{v_t}}
\end{equation}
where $\boldsymbol{{v_t}} \sim \mathcal{N}\left( {0,\boldsymbol{R_x}} \right)$ is measurement noise with covariance $\boldsymbol{R_x}$.

To solve this recursive estimation problem, we adopt the EKF due to its balance between computational efficiency and modeling flexibility. The EKF linearizes the state transition and observation models locally around current estimates and performs the following standard prediction and update steps:

1) State Prediction:
\begin{equation}\label{ekf1}
\boldsymbol{{\hat{x}_{t|t - 1}}} = \boldsymbol{{\hat{x}_{t-1 |t - 1}}}.
\end{equation}

2) Calculate Prediction Covariance Matrix:
\begin{equation}\label{ekf2}
\boldsymbol{{P_{t|t-1}}} = \boldsymbol{{P_{t-1|t-1}}} + \boldsymbol{Q_x}.
\end{equation}

3) Calculate Filter Gain:
\begin{equation}\label{ekf3}
\boldsymbol{{K_t}} = \boldsymbol{{P_{t-1|t-1}}}{\left( {\boldsymbol{{P_{t-1|t-1}}}+\boldsymbol{R_x}} \right)^{ - 1}}.
\end{equation}

4) Update States:
\begin{equation}\label{ekf4}
\boldsymbol{{\hat{x}_{t|t }}} = \boldsymbol{{\hat{x}_{t|t - 1}}} + \boldsymbol{{K_t}} \left( {\boldsymbol{{z_t}} - \boldsymbol{{\hat{x}_{t|t - 1}}}} \right).
\end{equation}

5) Update Prediction Covariance Matrix:
\begin{equation}\label{ekf5}
\boldsymbol{{P_{t|t}}} = \left(\boldsymbol{\rm I}-\boldsymbol{{K_t}}\boldsymbol{{H_t}}\right)\boldsymbol{{P_{t|t-1}}}.
\end{equation}

The use of the EKF is particularly well-justified in this context. First, the TO and the CFO follow a smooth first-order Markov process, a common assumption for oscillator-driven clock drift. Second, the LoS-based residuals ${\tau _{\rm{off}}\left( t \right)}$ and ${f _{\rm{off}}\left( t \right)}$
behave as approximately linear observations of the underlying offsets, especially in short intervals. Third, the EKF has been widely applied in OFDM synchronization and radar state estimation due to its ability to handle noisy, time-varying systems efficiently \cite{e1,e2}.

By integrating an EKF into the TVOE pipeline, we obtain real-time, drift-aware tracking of TO and CFO, forming the basis for a subsequent compensation of both LoS and NLoS components.
\subsection{Offset Compensation for NLoS Paths}
We employ CSI estimation to extract the sensing information, which can be estimated using methods such as the Least Squares (LS) approach with training sequences. The estimated CSI at the $n$-th subcarrier of the $m$-th packet can be expressed as
\begin{equation}\label{esta}
\begin{array}{l}
\begin{aligned}
&\boldsymbol{{\hat{h}_{n,m}}} = \boldsymbol{{h_{n,m}}} + \boldsymbol{z_{n,m,h}}\\
 &= \sum\limits_{l = 0}^L {{\beta _{l}}{e^{j2\pi m{T_s} {{f_{l}^e} } }} \times } {e^{ - j2\pi n\Delta f {{\tau _{l}^e}} }}   {\chi _{t,k}}\boldsymbol{{\alpha}\left( {q_{r,l}\left( t \right)} \right)}
 + \boldsymbol{z_{n,m,h}} ,
 \end{aligned}
\end{array}
\end{equation}
where $\boldsymbol{{h_{n,m}}} \in {\mathbb{C}^{{M_x^b  M_y^b }  \times  1 }}$, ${\chi _{t,k}} = \boldsymbol{{\alpha^T}\left( {q_{t,l}\left( t \right)} \right){w_t}}$ denotes the transmit beamforming gain, ${f_{l}^e = f_{l}\left( t \right) + f_{\rm{off}}\left( t \right)}$ and ${\tau _{l}^e}= {\tau _{l}\left( t \right) + \tau _{\rm{off}}\left( t \right)}$ denote the Doppler frequency and delay of $l$-th path, respectively; and $\boldsymbol{z_{n,m,h}}$ denotes the Gaussian noise vector with variance $\sigma _h^2$. Moreover, the parameters $f_{l}\left( t \right)$, $f_{\rm{off}}\left( t \right)$, $\tau _{l}\left( t \right)$, and $\tau _{\rm{off}}\left( t \right)$ are the real Doppler frequency, CFO, the real delay, and the TO of the $l$-th path, respectively. Due to the effects of TO and CFO, precise sensing information cannot be obtained directly at the receiver. Therefore, we acquire the corresponding information during the channel estimation process.
By stacking all $M_s \times N$ CSI estimates, the estimated correlation matrix of $\boldsymbol{{\boldsymbol{\hat{H}}}} \in {\mathbb{C}^{{M_x^b  M_y^b }  \times  {M_s}N }}$ is written as
\begin{equation}
\boldsymbol{{R_{h}}} = {{{\boldsymbol{\hat{H}}}\boldsymbol{\hat{H}^H}} \mathord{\left/
 {\vphantom {h {\left( {M \times N} \right)}}} \right.
 \kern-\nulldelimiterspace} {\left( {M_s N} \right)}}  ,
\end{equation}
where
\begin{equation}
{{\boldsymbol{\hat{H}}}\boldsymbol{\hat{H}^H}} = \sum\limits_{n = 1}^N {\sum\limits_{m = 1}^{{M_s}} {{\boldsymbol{{\hat{h}_{n,m}}}\boldsymbol{{\hat{h}_{n,m}^H}}} }  } .
\end{equation}

Obviously, the channel correlation matrix does not contain the phase offset caused by the TOs and CFOs. Thus, the AoA estimation will not be affected by the TOs and CFOs. Since the multipath signals are coherent, $M_s \times N$ CSI estimates are obtained within a CPI, $\boldsymbol{{R_{h}}}$ is rank deficient, i.e., $\rm{rank}(\boldsymbol{{R_{h}}})=1$, which degrades the estimation performance of conventional MUltiple SIgnal Classification (MUSIC) severely. In order to provide a high resolution for closely spaced signals, we employ spatial-smoothing MUSIC to achieve super-resolution in bistatic sensing environments.

\begin{figure}[!t]
	\centering
	\includegraphics[width=0.25\textheight]{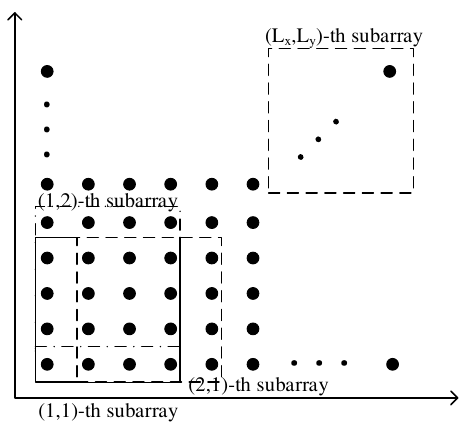}
	\DeclareGraphicsExtensions.
	\caption{The diagram of spatial smoothing.}
	\label{ss}
\end{figure}

To implement spatial-smoothing MUSIC, we divide the UPA composed of $M_x^b \times M_y^b$ receiving antennas into multiple overlapping subarrays as shown in Fig. \ref{ss}. Correspondingly, the $M_x^b$ array elements in the x-direction can be divided into $L_x$ interleaved sub-arrays and the $M_y^b$ array elements in the y-direction can be divided into $L_y$ interleaved sub-arrays. In this paper, we assume that $M_x^b = M_y^b$ and $L_x = L_y$. Then, let ${N_r^b} = {M_x^b-L_x+1} $, by taking advantage of the translation properties of a UPA, the original array can be translated into ${L_x} \times {L_x}$ copies of the sub-array whose sizes are $ {N_r^b}  \times  {N_r^b} $. Therefore, the covariance matrix of the $({l_x},{l_y})$-th subarray can be represented as
\begin{equation}
\boldsymbol{{R_h^{{l_x}{l_y}}}} = \boldsymbol{{T_{{l_x}{l_y}}}{R_h}}{\left(\boldsymbol{{T_{{l_x}{l_y}}}} \right)^H},
\end{equation}
where ${l_x} \in \left[ {1,2, \cdots ,{L_x}} \right]$, ${l_y} \in \left[ {1,2, \cdots ,{L_y}} \right]$, $\boldsymbol{{T_{{l_x}{l_y}}}} = \boldsymbol{{Z_y}} \otimes \boldsymbol{{Z_x}}$, and
\begin{equation}
\begin{array}{l}
\boldsymbol{{Z_x}} = \left[ {\left. {{\boldsymbol{0}_{{N_r^b} \times \left( {{l_x} - 1} \right)}}} \right|\left. {{\boldsymbol{I}_{N_r^b}}} \right|{\boldsymbol{0}_{{N_r^b} \times \left( {{L_x} - {l_x}} \right)}}} \right] \\
\boldsymbol{{Z_y}} = \left[ {\left. {{\boldsymbol{0}_{{N_r^b} \times \left( {{l_y} - 1} \right)}}} \right|\left. {{\boldsymbol{I}_{N_r^b}}} \right|{\boldsymbol{0}_{{N_r^b} \times \left( {{L_y} - {l_y}} \right)}}} \right].
\end{array}
\end{equation}
In this paper, ${\boldsymbol{0}_{x \times y}}$ and ${\boldsymbol{I}_{x }}$ denote the zero matrix with $x$ rows and $y$ columns and the identity matrix of size $x \times x$, respectively.

After spatial smoothing, the covariance matrix is written as
\begin{equation}
\boldsymbol{{\tilde {R}_h}} = \frac{1}{{{L_x}{L_y}}}\sum\limits_{{l_x} = 1}^{{L_x}} {\sum\limits_{{l_y} = 1}^{{L_y}} \boldsymbol{{{R_h^{{l_x}{l_y}}}}} }
\end{equation}
In short, the correlation of multipath signals can be reduced by forming subarrays and using the smoothed covariance matrix as input to the MUSIC algorithm.

By employing an EigenValue Decomposition (EVD) operation, the EVD of the sample covariance matrix $\boldsymbol{{\tilde {R}_h}}$ is denoted by
\begin{equation}\label{eigenA}
\left[ {\boldsymbol{{Q_h}},\boldsymbol{{\Lambda _h}}} \right] = {\rm{eig}}\left( {\boldsymbol{{\tilde {R}_h}}} \right),
\end{equation}
where $\boldsymbol{{\Lambda _h}}$ denotes the real-value diagonal matrix of eigenvalues in descending order and $\boldsymbol{{Q_h}}$ denotes the corresponding eigenvectors. Without additive noise, it has ${K}_s$ strongest positive real eigenvalues which correspond to the number of targets, and $N_r^u N_r^u-{K}_s$ small eigenvalues which correspond to the noise. The 2D MUSIC spatial spectrum is constructed as
\begin{equation}
{P_{2D}}\left( {{\varphi _l },{\theta _l}} \right) = \frac{1}{{{\boldsymbol{{\alpha ^H \left( {q_l} \left( t \right) \right)}}}\boldsymbol{{U_N}U_N^H}{\boldsymbol{{\alpha \left( {q_l}\left( t \right) \right)}}}}},
\end{equation}
where the columns of $\boldsymbol{{{U_N}}}$ span the noise subspace. To identify the AoAs of each target, the maxima of ${P_{2D}}\left( {{\varphi _l},{\theta _l}} \right)$ have to be found.

In order to estimate the delay and Doppler frequency information contained in the echo signal, we separate the multipath signals by using beamforming with a designed receive beamformer. Assuming that the AoAs of the multipaths are estimated accurately, we employ the LS method to design the beamforming vector corresponding to the $l$-th AoA. Using the desired array response matrix at ${K}_s$ directions $\boldsymbol{v} = \left[ {{v_1},{v_2}, \cdots ,{v_{{K_s}}}} \right] ^T $, we can compute the LS solution for a conventional beamforming design problem $\boldsymbol{A_qw} = \boldsymbol{v}$ as $\boldsymbol{{w_{LS}}} = {{\boldsymbol{{A_q}}}^\dag }\boldsymbol{v}$, where ${ \boldsymbol{A_q} ^\dag }$ denotes the pseudo-inverse of the actual array response matrix $\boldsymbol{A_q} = \left[ {\boldsymbol{\alpha \left( {{q_1}} \right)},\boldsymbol{\alpha \left( {{q_2}} \right)}, \cdots ,\boldsymbol{\alpha \left( {{q_{{K_s}}}} \right)}} \right]^T $ \cite{SM4}.

After employing the designed receive beamformer to separate the multipath signals, we can obtain
\begin{equation}\label{multi}
\begin{array}{l}
\begin{aligned}
{\hat{h}_{n,m}^l}=& \boldsymbol{{w_{LS}^H}}\boldsymbol{{\hat {h}_{n,m}}}\\
=& {h_{n,m}^l} + z_{n,m,h}^{l} \\
=& \sum\limits_{k \in {\Theta ^l}} {{\beta _{k}}{e^{j2\pi m{T_s}{f_k^e} }}} \times  {e^{ - j2\pi n\Delta f {{\tau _k^e}} }}{\chi _{t,k}}{\chi _{r,l,k}} + z_{n,m,h}^{l},
 \end{aligned}
\end{array}
\end{equation}
where ${h_{n,m}^l}$ denotes the actual channel response of $l$-th AoA, ${{\Theta ^l}}$ denotes the set of targets located in the $l$-th AoA, ${\chi _{r,l,k}}$ denotes the receive beamforming gain of $l$-th AoA and $z_{n,m,h}^{l}$ denotes the corresponding Gaussian noise vector. After stacking all $M_s \times N$ CSI estimates, the $l$-th channel response matrix is written as
$\left[ {\boldsymbol{{\hat{H}^l}}} \right]_{n,m} ={\hat{h}_{n,m}^l}$.

Then the delay and Doppler frequency of the targets can be extrapolated via a 2D MUSIC method. First, rewrite the $\left[ {\boldsymbol{{\hat{H}^l}}} \right]_{n,m}$ according to (\ref{multi}) as
\begin{equation}
\left[ {\boldsymbol{{\hat{H}^l}}} \right] =  \boldsymbol{{\alpha _{\tau ,l}}}{G_l}{\left( {\boldsymbol{{\alpha _{f,l}}}} \right)^T} + \boldsymbol{Z_h^{l}},
\end{equation}
where $\boldsymbol{{\alpha _{\tau ,l ,n}}} = {\left. {\left[ {{e^{ - j2\pi n\Delta f {{\tau _l^e} } }}} \right]} \right|_{n = 0,1, \cdots ,{N} - 1}} \in \mathbb{C}^{N \times 1} $, $\boldsymbol{{\alpha _{f,l,m}}} = {\left. {\left[ {{e^{j2\pi m{T_s} {{f_l^e} } }}} \right]} \right|_{m = 0,1, \cdots ,{M_s} - 1}} \in \mathbb{C}^{M_s\times 1}$, ${G_l} = {\beta _{c,l}}{\chi _{t,l}}{\chi _{r,l}}$ and $\boldsymbol{{\left[ {{Z_h^{l}}} \right]_{n,m}}}=\boldsymbol{z_{n,m,h}^{l}}$. Via a vectorization, the channel response of the $l$-th path is rewritten as
\begin{equation}
\boldsymbol{{\hat{h}^l}} = vec\left( {{\boldsymbol{{\hat{H}^l}}} } \right) = {G_l} \boldsymbol{{a_{f,l}}} \otimes \boldsymbol{{a_{\tau ,l}}} + vec\left( {\boldsymbol{Z_h^{l}} } \right).
\end{equation}

Moreover, the covariance matrix of $\boldsymbol{{\hat{h}_l}}$ is written as
\begin{equation}
\boldsymbol{R_{h}^l} = \frac{1}{{{M_s}{N}}}\sum\limits_{n = 1}^{{N}} {\sum\limits_{m = 1}^{{M_s}} \boldsymbol{{\hat{h}_{n,m}^l{{\left( {\hat{h}_{n,m}^l} \right)}^H}}} }.
\end{equation}
Here, we employ a 2D MUSIC based method to perform the joint estimation of the range and the Doppler frequencies. After employing the EVD operation, the covariance matrix can be decomposed as
\begin{equation}
\left[ {\boldsymbol{{Q_{h}^l}},\boldsymbol{{\Lambda _{h}^l}}} \right] = {\rm{eig}}\left( {\boldsymbol{{R_{h}^l}}} \right),
\end{equation}
where $\boldsymbol{{\Lambda _{h}^l}}$ denotes the real-value diagonal matrix of eigenvalues in descending order and $\boldsymbol{{Q_{h}^l}}$ denotes the corresponding eigenvectors. Then the spatial spectrum can be formulated as
\begin{equation}
P\left( {{\tau _l},{f_l}} \right) = \frac{1}{ {\left( \boldsymbol{{a_{f,l}}} \otimes \boldsymbol{{a_{\tau ,l}}} \right)^H \boldsymbol{U_{\tau ,f}^l{{\left( {U_{\tau ,f}^l} \right)}^H}}\left( \boldsymbol{{a_{f,l}}} \otimes \boldsymbol{{a_{\tau ,l}}} \right)}}
\end{equation}
where the columns of $\boldsymbol{U_{\tau ,f}^l}$ span the noise subspace. Then the peaks of the spatial spectrum correspond to the estimated $\hat{\tau_l}\left( t \right)$ and $\hat{f_l}\left( t \right)$, which can be searched by using the reduced-dimensional search method mentioned above.

\begin{figure}[!t]
	\centering
	\includegraphics[width=0.35\textheight]{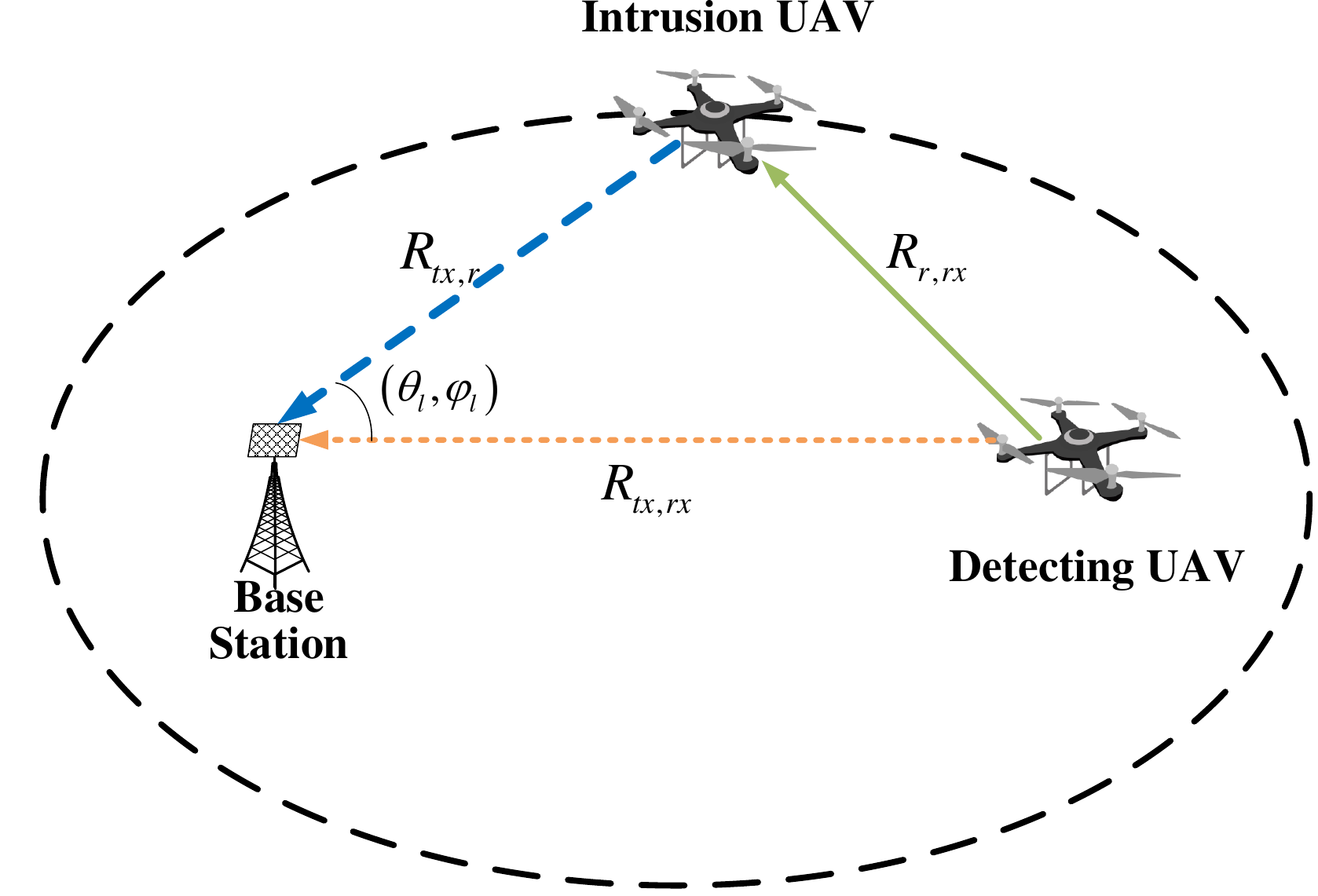}
	\DeclareGraphicsExtensions.
	\caption{The diagram of triangular location.}
	\label{locate}
\end{figure}

As shown in Fig.~\ref{locate}, the ISAC signal transmitted by the detecting UAV reaches the BS through both the LoS path and the NLoS path reflected by the intrusion UAV. By applying the TO and CFO deviations to the signal processing of the sensing targets' reflection path, we can estimate the actual signal propagation delay and Doppler frequency of the sensing targets' for the reflection path. Assuming that the estimated delay and Doppler frequency of $l$-th NLoS path are ${{\tau} _{l}^e}\left( t \right)$ and ${{f} _{l}^e}\left( t \right)$, respectively, the actual delay and Doppler frequency can be estimated as ${\tau _{l}}\left( t \right) = {{\tau} _{l}^e\left( t \right)} - {\hat{\tau} _{\rm{off}}\left( t \right)}$ and ${f _{l}\left( t \right)} = {{f} _{l}^e\left( t \right)} - {\hat{f} _{\rm{off}}\left( t \right)}$. After obtaining the accurate delay for the $l$-th NLoS path, the propagation range from the detecting UAV to the intrusion UAV, and then to the BS, is determined by ${R_l} = {R_{tx,r}} + {R_{r,rx}} = c{\tau _l}$, where $c$ denote the speed of light. Consequently, the intrusion UAV can be located on an ellipse with the BS and the detecting UAV as the two focuses. According to the estimation of Doppler frequency for the $l$-th NLoS path, the velocity of the intrusion UAV can be estimated accurately. Additionally, by estimating the direction of arrival $\left( {{\theta _l\left( t \right)},{\varphi _l\left( t \right)}} \right)$, the location of intrusion UAV can be uniquely determined on the ellipse.

\subsection{Summary}

\begin{algorithm}[!ht]
\caption{Time-Varying Offset Estimation Algorithm}\label{TVOE}
\begin{algorithmic}[1]
\STATE \textbf{Input}: CSI $\boldsymbol{{\hat{h}_{n,m}}}$, geometric delay $\tau _{\rm LoS}^{\rm geo} \left( t \right)$, Doppler $f _{\rm LoS}^{\rm geo} \left( t \right)$, EKF parameters $\boldsymbol{{Q_x}}$, $\boldsymbol{{R_x}}$.
\STATE \textbf{Output}: Estimated offsets ${\hat{\tau} _{\rm{off}}\left( t \right)}$, ${\hat{f} _{\rm{off}}\left( t \right)}$, precise sensing parameters ${\tau _{l}}\left( t \right)$, ${f _{l}}\left( t \right)$.
\STATE \textbf{Initialize}: EKF state $\boldsymbol{{\hat{x}_{0|0}}}$, $\boldsymbol{{{P}_{0|0}}}$.
\STATE \textbf{Step 1: LoS-based residual construction}
\STATE Compute delay and Doppler residuals $\tau _{\rm off} \left( t \right)$ and $f _{\rm off} \left( t \right)$.
\STATE \textbf{Step 2: EKF prediction and update}
\STATE $\boldsymbol{{\hat{x}_{t|t - 1}}} = \boldsymbol{{\hat{x}_{t-1 |t - 1}}}$.
\STATE $\boldsymbol{{\hat{x}_{t|t }}} = \boldsymbol{{\hat{x}_{t|t - 1}}} + \boldsymbol{{K_t}} \left( {\boldsymbol{{z_t}} - \boldsymbol{{\hat{x}_{t|t - 1}}}} \right)$.
\STATE \textbf{Step 3: Offset compensation}
\STATE ${\tau _{l}}\left( t \right) = {{\tau} _{l}^e\left( t \right)} - {\hat{\tau} _{\rm{off}}\left( t \right)}$ and ${f _{l}\left( t \right)} = {{f} _{l}^e\left( t \right)} - {\hat{f} _{\rm{off}}\left( t \right)}$.
\STATE \textbf{return estimated offsets $\boldsymbol{{\hat{x}_{t}}}$ and precise sensing parameters ${\tau _{l}}\left( t \right)$, ${f _{l}}\left( t \right)$.}
\end{algorithmic}
\end{algorithm}

As shown in Algorithm~\ref{TVOE}, the proposed TVOE algorithm provides a robust synchronization solution for UAV-enabled bistatic ISAC systems, where the transmitter and the receiver operate asynchronously due to independent clocks and spatial separation. By treating the LoS path as a dynamic synchronization reference, the TVOE leverages its geometric predictability to extract the delay and the Doppler residuals, which implicitly carry the TO and the CFO information.

The TVOE framework consists of three tightly coupled components:

1) LoS-Based Residual Construction: Extracts the measured delay/Doppler of the LoS path and compares it with geometrically predicted values to isolate the TO/CFO-induced deviations.

2) Sequential Offset Estimation via EKF: Models offset evolution as a stochastic process and applies sequential Bayesian filtering (i.e., EKF) to track the TO and the CFO over time, capturing drift and dynamics in real-time.

3) Offset Compensation for NLoS Paths: Applies the estimated offsets to all received paths, including weaker NLoS components, enabling accurate range, Doppler, and angle estimation for targets of interest under asynchronous conditions.

Unlike traditional synchronization schemes that rely on static references, global timing pulses, or tight hardware-level locking, the TVOE is fully software-defined, robust to the UAV mobility, and compatible with infrastructure-less or adversarial environments. It enables real-time offset tracking, improves range-Doppler resolution, and enhances both communication and sensing performance under dynamic deployment scenarios.


\section{Performance Analysis}
In this section, we conduct a performance analysis to evaluate the effectiveness of the proposed method in terms of the estimation of the number of targets, the Cramer-Rao Bound (CRB) of the estimated parameters, and the maximum sensing range and blind zone.

\subsection{Estimation of the Number of Targets}
To verify the availability of bistatic sensing, we estimate the number of targets to ensure it can sense all the targets. Actually, the number of targets (identifiable AoAs in the deployed scenario) can be estimated by model order selection with the basis of information theoretic criteria. According to (\ref{eigenA}), the covariance matrix $\boldsymbol{\tilde {R}_h}$ can be decomposed by performing an eigenvalue decomposition, where $\boldsymbol{{\Lambda _h}} = {\rm{diag}}\left( {{\lambda _1},{\lambda _2}, \cdots ,{\lambda _{K_s }}} \right)$ denotes the real-valued diagonal matrix of eigenvalues in descending order and $\boldsymbol{{Q_h}} \in {\mathbb{C}^{{K_s }  \times  {K_s } }}$ denotes the corresponding eigenvectors.

After employing the Minimum Description Length (MDL) criterion, the estimated number of targets is
\begin{equation}
{\hat{K_s}} = \mathop {\arg \min }\limits_{s \in \left\{ {0,1, \cdots ,{N_r^u } - 1} \right\}} \left\{ {{\rm{MDL}}\left( s \right)} \right\},
\end{equation}
where
\begin{equation}
\begin{array}{l}
\begin{aligned}
{\rm{MDL}}\left( s \right) = & - \ln {\left( {\frac{{\prod\nolimits_{i = s + 1}^{{N_r^u }} {\lambda _i^{{1 \mathord{\left/
 {\vphantom {1 {\left( {{N_r^u } - s} \right)}}} \right.
 \kern-\nulldelimiterspace} {\left( {{N_r^u } - s} \right)}}}} }}{{\frac{1}{{{N_r^u } - s}}\sum\nolimits_{i = s + 1}^{{N_r^u }} {{\lambda _i}} }}} \right)^{\left( {{N_r^u } - s} \right)M_s N}} \\
 &+ \frac{1}{2}s\left( {2{N_r^u } - s} \right)\ln \left( {M_s N} \right).
 \end{aligned}
\end{array}
\end{equation}

\subsection{CRB of Estimated Parameters}
After obtaining the estimated parameters of AoA, delay and Doppler frequency, we derive the CRB, which can provide a benchmark for evaluating the accuracy of the estimated parameters. According to (\ref{esta}), the ideal CSI expression for estimating the AoAs of multipath signals is
\begin{equation}\label{crba1}
\boldsymbol{{\hat{h}_{n,m}}} = \sum\limits_{l = 0}^L {\boldsymbol{{b_{n,m,l}}}}  + \boldsymbol{z_{n,m,h}},
\end{equation}
where $\boldsymbol{{b_{n,m,l}}} = {{\beta _{c,l}}{e^{j2\pi m{T_s} {{f_{c,l}^e}} }}  } {e^{ - j2\pi n\Delta f {{\tau _{c,l}^e}} }} {\chi _{t,k}}\boldsymbol{\alpha \left( {q_l} \right)} $. Since $\boldsymbol{z_{n,m,h}}$ follows a Gaussian distribution, $\boldsymbol{{\hat{h}_{n,m}}}$ follows a Gaussian distribution with mean value $\sum\limits_{l = 0}^L {\boldsymbol{{b_{n,m,l}}}}$ and variance $\sigma _h^2$. Thus, the Probability Density Function (PDF) can be expressed as
\begin{equation}\label{crba2}
p_1\left( \boldsymbol{h} \right) = \frac{1}{{\pi \sigma _h^2}}{e^{ - \frac{1}{{\sigma _h^2}}{{\left| {\boldsymbol{h} - \sum\limits_{l = 0}^L {\boldsymbol{{b_{n,m,l}}}}} \right|}^2}}}.
\end{equation}

The CRB provides a lower limit on the variance obtainable by any technique as a function of the Fisher Information Matrix (FIM) and the estimator's bias gradient \cite{crb}. The CRB of AoA estimation is shown as follows, the detailed derivation is demonstrated in the Appendix.

\begin{equation}
\begin{array}{l}
\begin{aligned}
{\rm{CRB}}\left( {{\varphi _l}} \right) = & {\left[ {{{\left( {{M_s}N\boldsymbol{F\left( {\varphi _l, \theta _l} \right)}} \right)}^{ - 1}}} \right]_{l,l}}, \\
= & \frac{1}{{{M_s}N}}{\left[ {{{\left( \boldsymbol{{F\left( {\varphi _l, \theta _l} \right)}} \right)}^{ - 1}}} \right]_{l,l}}.
\end{aligned}
\end{array}
\end{equation}

\begin{equation}
{\rm{CRB}}\left( {{\theta _l}} \right) = \frac{1}{{M_sN}}{\left[ {{{\left( \boldsymbol{{F\left( {{\varphi _l},{\theta _l}} \right)}} \right)}^{ - 1}}} \right]_{l + K_s \times 1,l + K_s \times 1}}.
\end{equation}

After estimating the AoAs, the multipath signals can be separated. According to (\ref{multi}), the ideal CSI expression for estimating the delay and Doppler frequency are
\begin{equation}
\hat{h}_{n,m}^l = {d_{n,m,l}} + z_{n,m,h}^{l},
\end{equation}
where ${D_{n,m,l}} = {{\beta _{c,l}}{e^{j2\pi m{T_s} {{f_{c,l}^e}} }}  } {e^{ - j2\pi n\Delta f {{\tau _{c,l}^e}} }} {\chi _{t,k}}{\chi _{r,l,k}} $. Thus, the PDF can be expressed as
\begin{equation}
p_2\left( h_2 \right) = \frac{1}{{\pi \sigma _{h,l}^2}}{e^{ - \frac{1}{{\sigma _{h,l}^2}}{{\left| {h_2 - {{d_{n,m,l}}}} \right|}^2}}},
\end{equation}

Similarly, the CRB of the delay and Doppler frequency are obtained as
\begin{equation}
\begin{array}{l}
\begin{aligned}
{\rm{CRB}}\left( {{\tau _l}} \right) &= {\left[ {{{\left( {M_sN\boldsymbol{F\left( {{\tau_l },{f_l}} \right)}} \right)}^{ - 1}}} \right]_{l,l}}\\
 &= \frac{1}{{M_sN}}{\left[ {{{\left( \boldsymbol{{F\left( {{\tau _l},{f_l}} \right)}} \right)}^{ - 1}}} \right]_{l,l}},
\end{aligned}
\end{array}
\end{equation}

\begin{equation}
{\rm{CRB}}\left( {{f _l}} \right) = \frac{1}{{M_sN}}{\left[ {{{\left( \boldsymbol{{F\left( {{\tau _l},{f_l}} \right)}} \right)}^{ - 1}}} \right]_{l + K_s \times 1,l + K_s \times 1}}.
\end{equation}

\subsection{Maximum Sensing Range and Blind Zone}
In the bistatic sensing configuration, the range of targets can be estimated through echo processing, which is related to the distance between the transmitter and the target, ${R_{tx,r}}$, and that between the target and the receiver, ${R_{r,rx}}$, via ${R_l} = {R_{tx,r}} + {R_{r,rx}} = c{\tau _l}$. As depicted in Fig. \ref{locate}, after estimating the range, the target is located on an ellipse whose the focuses are the transmitter and the receiver. In order to obtain unambiguous range detection, Inter Symbol Interference (ISI) needs to be avoided, so that the guard interval $T_g$ should exceed the propagation delay. Then the maximum of the sensing range is ${R_l} \le {T_g}c + {R_{tx,rx}}$ in a bistatic sensing configuration while the minor axis of this maximum ellipse is ${A_{\max }} = \sqrt {{{\left( {{T_g}c + {R_{tx,rx}}} \right)}^2} - R_{tx,rx}^2} $.

Notably, if the target lies in a region enclosing the line between the transmitter and the receiver, then it falls into the blind zone and may be hard to be detected due to the range resolution. Then the minimum of the sensing range is ${R_l} \ge {R_{tx,rx}} + \Delta r$, where $\Delta r = {c \mathord{\left/
 {\vphantom {c {N\Delta f}}} \right.
 \kern-\nulldelimiterspace} {N\Delta f}}$ denotes the resolution of range estimation. As the receiver cannot distinguish the reflected path from the direct one if the range is below ${R_{tx,rx}} + \Delta r$, then the minimum ellipse is determined, whose major axis is ${R_{tx,rx}} + \Delta r$ and minor axis is ${A_{\min }} = \sqrt {{{\left( {{R_{tx,rx}} + \Delta r} \right)}^2} - R_{tx,rx}^2} $.

 \begin{table}[!t]
	\centering
	\caption{Simulation Parameters \cite{int14,SM3,para1,MU1}}\label{simu}
	\begin{tabular}{c|c}
		\hline
		\hline
		\label{Parameter:simulation}
		{\textbf{Parameter}}  & {\textbf{Value}} \\
		\hline
		Total transmit power (${P_{\rm{total}}}$)  & 20 dBm \\
        Carrier frequency (${f_c}$) & 28 GHz\\
        Carrier wavelength ($\lambda $) & ${c \mathord{\left/
 {\vphantom {c {{f_c}}}} \right.
 \kern-\nulldelimiterspace} {{f_c}}}$ \\
        Bandwidth ($B$)  & 100 MHz \\
        OFDM packets number ($M_s$)  & 64 \\
        Subcarrier number ($N$) & 1024 \\
        Subcarrier interval ($\Delta f$) & ${B \mathord{\left/
 {\vphantom {B N}} \right.
 \kern-\nulldelimiterspace} N}$ \\
        Cyclic Prefix ($T_g$) & $3.34{e^{ - 6}} s$ \\
        NLoS path number ($L$)  & 4 \\
		Speed of light ($c$) & $3 \times {10^8}$ m/s \\
		\hline
		\hline
	\end{tabular}
\end{table}

\section{Simulation Results}
In this section, we present numerical results to validate the performance of our proposed LoS-aided offset mitigation method in terms of estimation accuracy for bistatic sensing. For this evaluation, we assume that the detecting UAV functions as an aerial base station, connected to a ground BS, and is capable of moving freely within an area of 1000m $\times$ 1000m based on control commands from the ground BS. The locations of the BS and the detecting UAV are set to (0, 0, 0) and (100, 100, 100), respectively. The intrusion UAVs are randomly distributed within the bistatic sensing range. Additionally, the AWGN power at each receiver is set to ${\sigma _n ^2} = -174$ dBm/Hz, assuming a total transmit power of $P_{\rm{total}} = 1$ W. The detailed simulation parameters are shown in Table~\ref{simu}.

\begin{figure}[!t]
	\centering
	\DeclareGraphicsExtensions.
    \subfigure[]{\includegraphics[width=0.35\textheight]{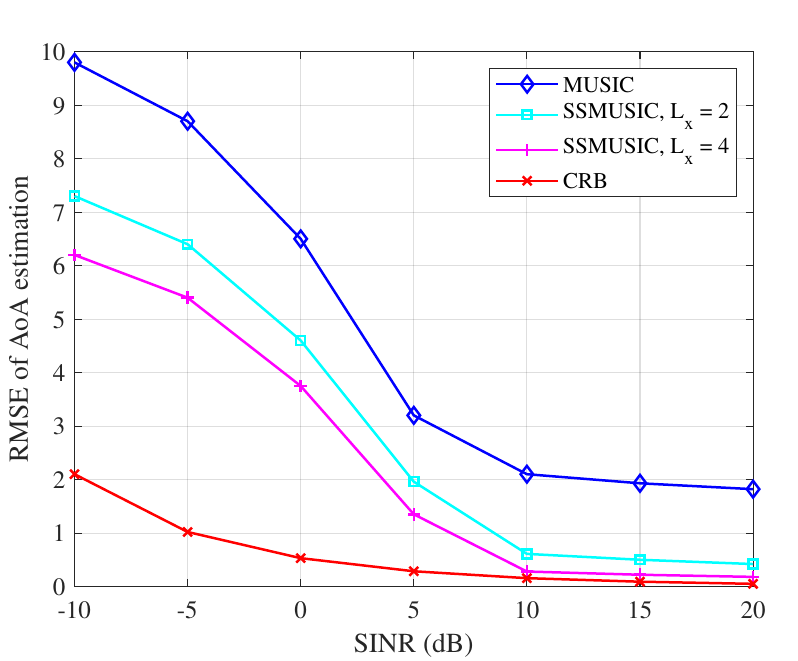}\label{RMSEAoA}}
    \hspace{0.5cm}
    \subfigure[]{\includegraphics[width=0.35\textheight]{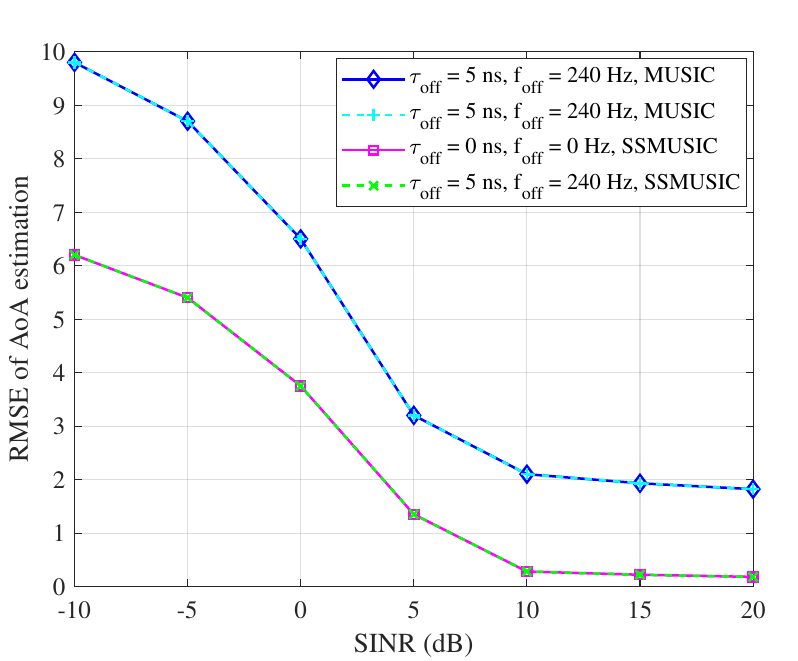}\label{RMSEAoA2}}
	\caption{(a) The RMSE of AoA estimation vs. SINR using the MUSIC and spatial smoothing MUSIC schemes. (b) The RMSE of AoA estimation vs. SINR under various TOs and CFOs.}
	\label{RMSEA}
\end{figure}

Note that the BS knows the locations of the detecting UAV in advance, and the detecting UAV flies according to the control command issued by the BS. After the detecting UAV transmits the signals, the signals for communication are received directly by the BS, while the signals for sensing are reflected off the surrounding targets for radar sensing. To evaluate the performance of the proposed LoS-aided offset mitigation method, we employ the Root Mean Squared Error (RMSE) as the estimation performance metric. The estimation Mean Squared Error (MSE) is defined as the average of all squared errors in the estimation results under a given set of simulation parameters, and the RMSE is the square root of the MSE.

First, we compare the proposed method with the traditional MUSIC algorithm in terms of the estimation accuracy of the AoAs. Fig.~\ref{RMSEA} presents the estimation RMSE of the AoAs versus the SINR of the sensing signals under different methods and various TOs and CFOs conditions. It is evident that, in the presence of perfectly correlated sources, such as during multipath propagation, the traditional MUSIC algorithm struggles to estimate the AoAs accurately due to the nonsingular nature of the measurement covariance matrix. The spatial smoothing technique, which employs a UPA with subarray averaging, improves the AoA estimation accuracy by effectively increasing the array's resolution and mitigating the effects of noise. Fig.~\ref{RMSEAoA} illustrates that spatial smoothing is highly effective in this condition, significantly improving angle estimation precision in a multipath environment with multiple targets, albeit at the cost of reduced array aperture. Notably, as the number of decomposed subarrays increases, the decoherence performance improves, with the estimation RMSE of the AoAs approaching the square root of its CRB. Compared to the traditional MUSIC algorithm, the proposed spatial smoothing MUSIC (SSMUSIC) algorithm enhances the RMSE performance by 85\%. Furthermore, Fig.~\ref{RMSEAoA2} demonstrates that, even under varying TOs and CFOs, the estimation RMSE of the AoAs remains consistent, unaffected by these offsets.

\begin{figure}[!t]
	\centering
	\DeclareGraphicsExtensions.
    \subfigure[]{\includegraphics[width=0.35\textheight]{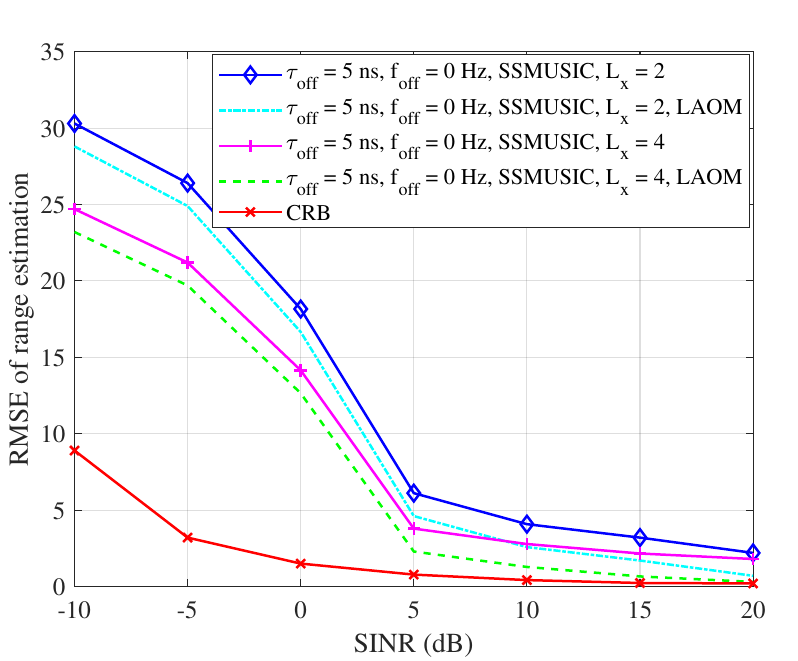}\label{RMSERan}}
    \hspace{0.5cm}
    \subfigure[]{\includegraphics[width=0.35\textheight]{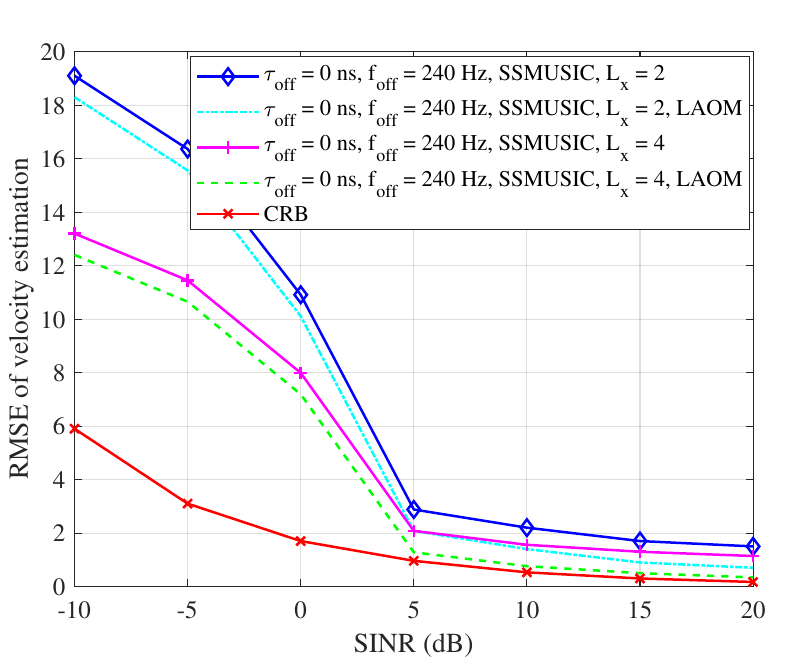}\label{RMSEVel}}
	\caption{(a) The RMSE of range estimation vs. SINR under various TOs. (b) The RMSE of velocity estimation vs. SINR under various CFOs.}
	\label{RMSER}
\end{figure}

Fig.~\ref{RMSER} illustrates the estimation RMSE of the range and velocity versus the SINR of the sensing signals under various TOs and CFOs conditions. In Fig.~\ref{RMSER}, TO and CFO are evaluated separately to isolate their individual impacts on sensing performance. Specifically, TO is varied while keeping $\rm{CFO} = 0$ in Fig.~\ref{RMSERan}, and CFO is varied with $\rm{TO} = 0$ in Fig.~\ref{RMSEVel}, allowing for clearer attribution of estimation error sources. It can be seen that the presence of TOs and CFOs introduces ambiguity in the estimation of both range and velocity. Fig.~\ref{RMSERan} presents the estimation RMSE of range under varying TOs. As more subarrays are decomposed, decoherence performance improves, allowing for better separation of multipath signals and more accurate velocity estimation. Our proposed LAOM method effectively eliminates the ambiguity in range estimation by utilizing the reference LoS path, achieving a performance close to the square root of the CRB. Fig.~\ref{RMSEVel} shows the estimation RMSE of velocity under varying CFOs. The LAOM method successfully mitigates the effects of CFOs, also approaching the square root of the CRB.


\begin{figure}[!t]
	\centering
	\includegraphics[width=0.35\textheight]{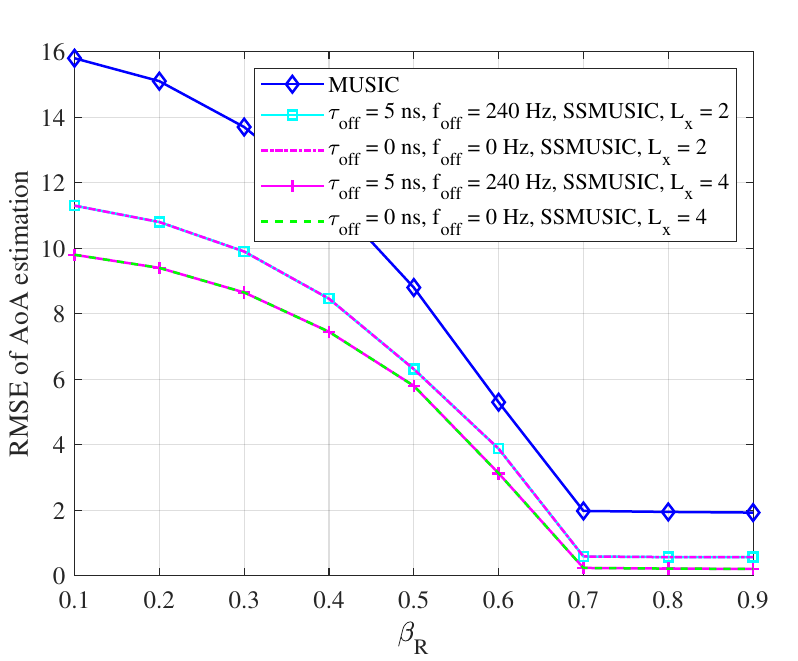}
	\DeclareGraphicsExtensions.
	\caption{The RMSE of AoA estimation vs. power distribution factor $\beta_R$.}
	\label{RMSEAoAp}
\end{figure}

Fig.~\ref{RMSEAoAp} depicts the estimation RMSE of the AoAs versus the power distribution factor $\beta_R$ after employing both the traditional MUSIC and spatial smoothing MUSIC algorithms. It is obvious that as the power distribution factor increases, the power allocated for sensing also increases, leading to a rapid decrease in the estimation RMSE of the AoAs, which eventually stabilizes as it reaches the performance limits under these conditions. Furthermore, the estimation performance of the AoAs primarily relies on the effectiveness of decoherence, which is not affected by TOs and CFOs.

\begin{figure}[!t]
	\centering
	\DeclareGraphicsExtensions.
    \subfigure[]{\includegraphics[width=0.35\textheight]{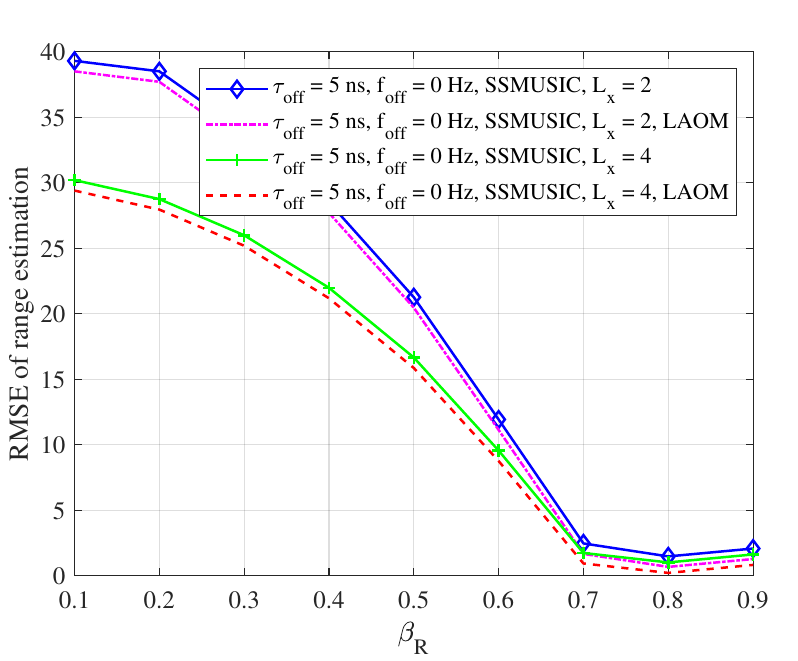}\label{RMSERanp}}
    \hspace{0.5cm}
    \subfigure[]{\includegraphics[width=0.35\textheight]{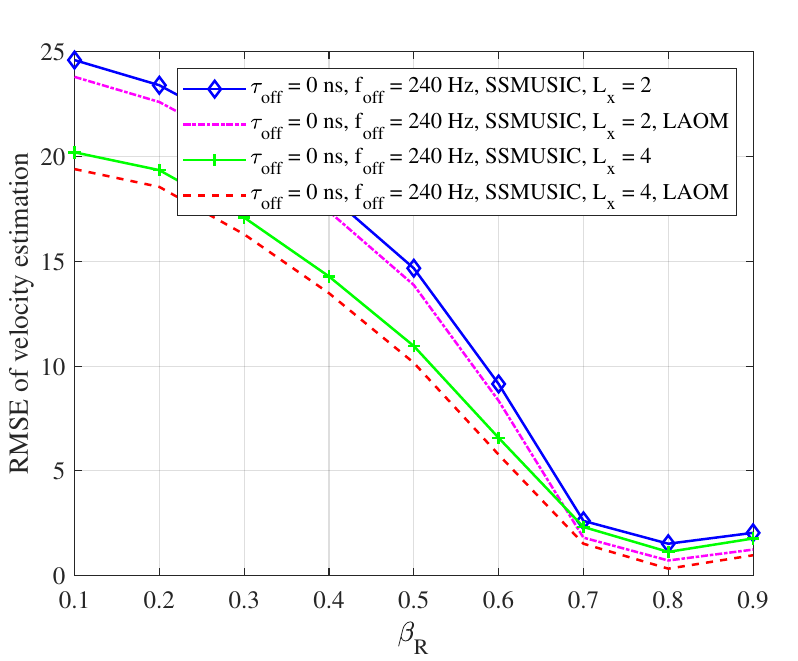}\label{RMSEVelp}}
	\caption{(a) The RMSE of range estimation vs. power distribution factor $\beta_R$ under various TOs. (b) The RMSE of velocity estimation vs. power distribution factor $\beta_R$ under various CFOs.}
	\label{RMSEV}
\end{figure}

Fig.~\ref{RMSEV} presents the estimation RMSE of the range and velocity versus the power distribution factor $\beta_R$ under various TOs and CFOs conditions. As the power distribution factor increases, the SINR for sensing improves as anticipated. However, if the power distribution factor becomes excessively large, communication performance may degrade due to a larger portion of resources being allocated to sensing, indicating a trade-off between communication and sensing functions. Fig.~\ref{RMSERanp} shows that while the estimation RMSE of range decreases rapidly initially with increasing power distribution factor, excessive allocation leads to communication degradation, resulting in a slight increase in RMSE. Fig.~\ref{RMSEVelp} illustrates the estimation performance of velocity under varying power distribution factor conditions. Given that the accuracy of range and velocity estimation is dependent on AoA estimation, our proposed LAOM method achieves a robust estimation performance with the aid of the spatial smoothing MUSIC algorithm.


\begin{figure}[!t]
	\centering
	\DeclareGraphicsExtensions.
    \subfigure[]{\includegraphics[width=0.35\textheight]{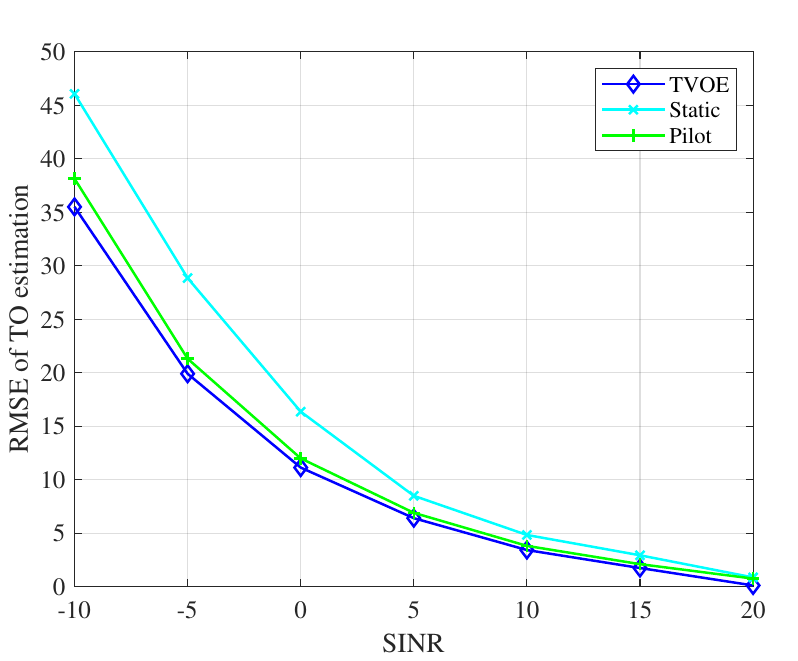}\label{RMSETO}}
    \hspace{0.5cm}
    \subfigure[]{\includegraphics[width=0.35\textheight]{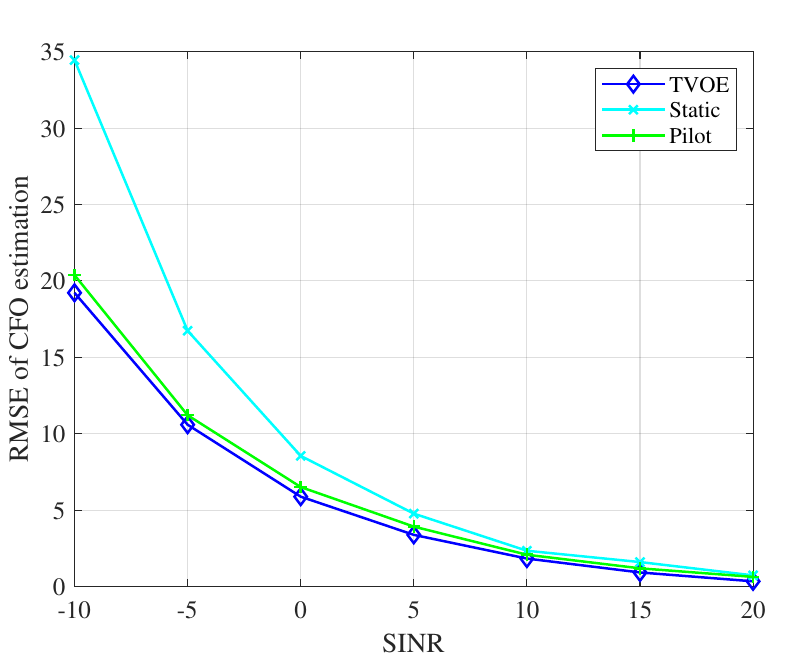}\label{RMSECFO}}
	\caption{(a) The RMSE of TO estimation vs. SINR using TVOE, static and pilot methods. (b) The RMSE of CFO estimation vs. SINR using TVOE, static and pilot methods.}
	\label{RMSET}
\end{figure}

Fig.~\ref{RMSET} illustrates the RMSE performance of the TO/CFO estimation as a function of the SINR for three representative schemes: the proposed TVOE method, a conventional static LoS-based estimator \cite{int15}, and a pilot-based estimator \cite{e1} that leverages known training sequences. For the TO estimation, the RMSE of the static method exceeds 40 ns at low SINR, while the TVOE maintains a robust performance below 25 ns due to its ability to model the temporal offset evolution via sequential filtering. At high SINRs, the TVOE achieves nearly 60\% lower TO RMSE than the static baseline. For the TO and the CFO estimation, the pilot-based estimator yields the best accuracy at high SINR, leveraging coherent averaging over known pilots. However, its reliance on pilot structures limits its applicability in passive sensing or infrastructure-free UAV scenarios. Notably, under low SINR, the pilot-based method's performance deteriorates rapidly due to pilot contamination, while the TVOE maintains reliable estimation thanks to its anchor-free operation using the LoS path as a dynamic reference. These results confirm that the TVOE enables reliable, real-time offset estimation without requiring pilot exchange or GNSS synchronization, rendering it particularly suitable for UAV-enabled bistatic ISAC systems operating under dynamic environments.

\section{Conclusion}
In this paper, we have addressed the fundamental challenge of clock asynchronism in UAV-enabled bistatic ISAC systems by introducing the TVOE framework. Unlike traditional methods that assume static channel conditions or require explicit pilot signaling, the TVOE is designed for fully dynamic and pilot-free environments, where synchronization must be achieved passively and adaptively. The core idea is to exploit the LoS path as a naturally observable synchronization reference, leveraging its geometric predictability to track timing and frequency offsets via a sequential filtering approach. By formulating a state-space model and applying the EKF, the framework enables continuous offset tracking without requiring shared clocks or GNSS alignment. Moreover, the TVOE corrects residual misalignments in NLoS components, thereby enhancing sensing fidelity even under mobility and oscillator drift.

Beyond the algorithmic development, our simulations demonstrate that TVOE provides significant performance gains over static and pilot-based baselines. Importantly, the proposed method maintains compatibility with standard OFDM waveforms, rendering it practical for real-world deployment in 6G ISAC systems. This study not only offers a technical solution to synchronization under dynamic conditions but also sets the foundation for a new class of self-referenced, mobility-aware synchronization mechanisms. Future research may extend this work to multi-target scenarios, incorporate learning-based state prediction, or generalize it to multi-node cooperative ISAC networks with decentralized architectures.

\appendix
\section*{Derivation of CRB}
According to (\ref{crba1}) and (\ref{crba2}), the log-likelihood function of $h$ is
\begin{equation}
L\left( \boldsymbol{h} \right) =  - NM_s\ln \left( {\pi \sigma _h^2} \right) - \frac{1}{{\sigma _h^2}}\sum\limits_{n,m}^{N,M_s} {{{\left| {{\boldsymbol{h}} - \sum\limits_{l = 0}^L {\boldsymbol{{b_{n,m,l}}}}} \right|}^2}}.
\end{equation}

Then, the FIM of AoA estimation can be given as
\begin{equation}
\boldsymbol{F\left( \varphi _l, \theta _l \right)} = - E {\left( {\frac{{\partial^2 L\left( \boldsymbol{h} \right)}}{{\partial^2 \left[\alpha \left( \boldsymbol{{q_{l}}} \right)\right]}}} \right)} =  - \left[ {\begin{array}{*{20}{c}}
{\boldsymbol{{E_{\varphi \varphi }}}}&{\boldsymbol{{E_{\varphi \theta }}}}\\
{\boldsymbol{{E_{\varphi \theta }}}}&{\boldsymbol{{E_{\theta \theta }}}}
\end{array}} \right].
\end{equation}

Since the $\boldsymbol{{q_l}}$ contains the azimuth angle $\varphi _l$ and elevation angle $\theta _l$ information, we need to derive the partial derivative of $L\left( \boldsymbol{h} \right)$ with respect to $\varphi _l$ firstly, which is computed as follows.
\begin{equation}
\begin{array}{l}
\begin{aligned}
\frac{{\partial L\left( \boldsymbol{h} \right)}}{{\partial \varphi _l}} & = \frac{1}{{\sigma _h^2}}\left[ {\sum\limits_{n,m}^{N,M_s} {\left( {{\boldsymbol{h}} - \sum\limits_{l = 0}^L {\boldsymbol{{b_{n,m,l}}}}} \right)\frac{{\partial {{\left( {\sum\limits_{l = 0}^L {\boldsymbol{{b_{n,m,l}}}}} \right)}^*}}}{{\partial \varphi _l}}} } \right.\\
& \left. { + \frac{{\partial \left(\sum\limits_{l = 0}^L {\boldsymbol{{b_{n,m,l}}}}\right)}}{{\partial \varphi _l}}\left[ {\boldsymbol{h}^ *  - \left(\sum\limits_{l = 0}^L {\boldsymbol{{b_{n,m,l}}}}\right)^ * } \right]} \right].
\end{aligned}
\end{array}
\end{equation}

Obviously,
\begin{equation}
\frac{{\partial {{\left( \sum\limits_{l = 0}^L {\boldsymbol{{b_{n,m,l}}}} \right)}^*}}}{{\partial \varphi _l}} = {\left[ {\frac{{\partial \left( \sum\limits_{l = 0}^L {\boldsymbol{{b_{n,m,l}}}} \right)}}{{\partial \varphi _l}}} \right]^*},
\end{equation}
\begin{equation}\label{part1}
\frac{{\partial {\left( \sum\limits_{l = 0}^L {\boldsymbol{{b_{n,m,l}}}} \right)}}}{{\partial \varphi _l}} = j\pi \boldsymbol{{A_{p,q}^ {1,l}}} \circ \boldsymbol{{b_{n,m,l}}},
\end{equation}
where $\boldsymbol{{A_{p,q}^ {1,l}} }= {\left. {\left[ {{a_{p,q}^ {1,l}}} \right]} \right|_{p = 0,1, \cdots ,M_y^u - 1,q = 0,1, \cdots ,M_x^u - 1}}$ and ${a_{p,q}^ {1,l}} = p\Omega _{l,y} - q\Omega _{l,x}$. Then the first-order partial derivative of $L\left( \boldsymbol{h} \right)$ with respect to $\varphi _l$ is obtained as
\begin{equation}\label{part11}
\frac{{\partial L\left( \boldsymbol{h} \right)}}{{\partial \varphi _l}} = \frac{1}{{\sigma _h^2}}\sum\limits_{n,m}^{N,M_s} {2{\mathop{\rm Re}\nolimits} \left\{ {j\pi \boldsymbol{{A_{p,q}^{1,l}}} \circ \boldsymbol{h}^ * \boldsymbol{{b_{n,m,l}}}} \right\}} .
\end{equation}

Similarly, the first-order partial derivative of $L\left( \boldsymbol{h} \right)$ with respect to $\theta _l$ is obtained as
\begin{equation}\label{part21}
\frac{{\partial L\left( \boldsymbol{h} \right)}}{{\partial \theta _l}} = \frac{1}{{\sigma _h^2}}\sum\limits_{n,m}^{N,M_s} {2{\mathop{\rm Re}\nolimits} \left\{ {j\pi \boldsymbol{{A_{p,q}^{2,l}}} \circ \boldsymbol{h}^ * \boldsymbol{{B_{n,m,l}}}} \right\}} .
\end{equation}
where $\boldsymbol{{A_{p,q}^ {2,l}}} = {\left. {\left[ {{a_{p,q}^ {2,l}}} \right]} \right|_{p = 0,1, \cdots ,M_y^u - 1,q = 0,1, \cdots ,M_x^u - 1}}$ and ${a_{p,q}^ {2,l}} = - p\cos {\theta _l}\cos {\varphi _l} - q\cos {\theta _l}\sin {\varphi _l}$.

According to (\ref{part1}) and (\ref{part11}), the second-order partial derivative ${\boldsymbol{{E_{\varphi \varphi }}}}$ with respect to $\varphi _l$ is obtained as
\begin{equation}
\begin{array}{l}
\begin{aligned}
& \frac{{{\partial ^2}L\left( \boldsymbol{h} \right)}}{{{\partial }\left(\varphi _l\right)^2}} = \frac{2 \pi}{{\sigma _h^2}}\sum\limits_{n,m}^{N,M_s} {{\mathop{\rm Re}\nolimits} \left[ { - 2\pi \boldsymbol{{A_{p,q}^{1,l} }} \circ } \boldsymbol{{A_{p,q}^{1,l}} } \circ {{\left| {\boldsymbol{{b_{n,m,l}}}} \right|}^2}+  \right.}\\
&\left. { j\boldsymbol{z_{n,m,h}^{l}} \times \left( {\left(\boldsymbol{A_{p,q}^{1,l}}\right)' \circ \boldsymbol{{b_{n,m,l}}} + j\pi \boldsymbol{{A_{p,q}^{1,l}} } \circ \boldsymbol{{A_{p,q}^{1,l} }} \circ \boldsymbol{{b_{n,m,l}}}} \right)} \right],
\end{aligned}
\end{array}
\end{equation}
where $\left(\boldsymbol{A_{p,q}^{1,l}}\right)'$ denotes the first-order derivative with respect to $\varphi _l$.

Moreover, if $i \ne j$, owing to the different path are independent, the ($i$,$j$)-th entry of ${\boldsymbol{{E_{\varphi \varphi }}}}$ is given by
\begin{equation}
\frac{{{\partial ^2}L\left( \boldsymbol{h} \right)}}{{\partial \varphi _{l}^{{i}}\partial \varphi _{l}^{{j}}}} = 0.
\end{equation}
Thus, the non-diagonal elements of ${\boldsymbol{{E_{\varphi \varphi }}}}$, ${\boldsymbol{{E_{\varphi \theta }}}}$ and ${\boldsymbol{{E_{\theta \theta }}}}$ are all zero.

Similarly, the second-order partial derivative ${\boldsymbol{{E_{\theta \theta }}}}$ with respect to $\theta _l$ is obtained as
\begin{equation}
\begin{array}{l}
\begin{aligned}
& \frac{{{\partial ^2}L\left( \boldsymbol{h} \right)}}{{{\partial }\left(\theta _l\right)^2}} = \frac{2 \pi}{{\sigma _h^2}}\sum\limits_{n,m}^{N,M_s} {{\mathop{\rm Re}\nolimits} \left[ { - 2\pi \boldsymbol{{A_{p,q}^{2,l} }} \circ } \boldsymbol{{A_{p,q}^{2,l} }} \circ {{\left| {\boldsymbol{{b_{n,m,l}}}} \right|}^2}+  \right.}\\
&\left. {j\boldsymbol{z_{n,m,h}^{l}} \times \left( {\left(\boldsymbol{A_{p,q}^{2,l}}\right)' \circ \boldsymbol{{b_{n,m,l}}} + j\pi \boldsymbol{{A_{p,q}^{2,l}} } \circ \boldsymbol{{A_{p,q}^{2,l}} } \circ \boldsymbol{{b_{n,m,l}}}} \right)} \right].
\end{aligned}
\end{array}
\end{equation}

Lastly, the second-order partial derivative ${\boldsymbol{{E_{\varphi \theta }}}}$ is obtained as
\begin{equation}
\begin{array}{l}
\begin{aligned}
& \frac{{{\partial ^2}L\left( \boldsymbol{h} \right)}}{{{\partial }\left(\varphi _l\right)^2}} = \frac{2 \pi}{{\sigma _h^2}}\sum\limits_{n,m}^{N,M} {{\mathop{\rm Re}\nolimits} \left[ { - 2\pi \boldsymbol{{A_{p,q}^{1,l}} } \circ } \boldsymbol{{A_{p,q}^{2,l}} } \circ {{\left| {\boldsymbol{{b_{n,m,l}}}} \right|}^2}+ \right.}\\
&\left. { j\boldsymbol{z_{n,m,h}^{l}} \times \left( {\left(\boldsymbol{A_{p,q}^{1,l}}\right)'' \circ \boldsymbol{{b_{n,m,l}}} + j\pi \boldsymbol{{A_{p,q}^{1,l}} } \circ \boldsymbol{{A_{p,q}^{2,l}} } \circ \boldsymbol{{b_{n,m,l}}}} \right)} \right],
\end{aligned}
\end{array}
\end{equation}
where $\left(\boldsymbol{A_{p,q}^{1,l}}\right)''$ denotes the second-order derivative with respect to $\theta _l$.


The detailed derivation process of delay and Doppler shift is similar to the one above. The FIM of delay and Doppler frequency estimation is given as
\begin{equation}
\boldsymbol{F\left( \tau , f \right) }= - E {\left( {\frac{{\partial^2 L\left( h_2 \right)}}{{  \partial \tau \partial f}}} \right)} =  - \left[ {\begin{array}{*{20}{c}}
{\boldsymbol{{E_{\tau \tau }}}}&{\boldsymbol{{E_{\tau f }}}}\\
{\boldsymbol{{E_{\tau f }}}}&{\boldsymbol{{E_{f f }}}}
\end{array}} \right],
\end{equation}
where
\begin{equation}
\begin{array}{l}
\begin{aligned}
&\frac{{{\partial ^2}L\left( h_2 \right)}}{{{\partial }\left(\tau _l\right)^2}} =\\
&-\frac{2 }{{\sigma _{h,l}^2}}{\left( {2\pi n\Delta f} \right)^2}\sum\limits_{n,m}^{N,M_s} {{\mathop{\rm Re}\nolimits} \left[ 2 {{\left| {{d_{n,m,l}}} \right|}^2}+ z_{n,m,h}^l {d_{n,m,l}} \right]},
\end{aligned}
\end{array}
\end{equation}
\begin{equation}
\begin{array}{l}
\begin{aligned}
&\frac{{{\partial ^2}L\left( h_2 \right)}}{{{\partial }\left(f _l\right)^2}} = \\
&-\frac{2 }{{\sigma _{h,l}^2}}{\left( {2\pi m T_s} \right)^2}\sum\limits_{n,m}^{N,M_s} {{\mathop{\rm Re}\nolimits} \left[ 2 {{\left| {{d_{n,m,l}}} \right|}^2}+ z_{n,m,h}^{l} {d_{n,m,l}} \right]},
\end{aligned}
\end{array}
\end{equation}
\begin{equation}
\begin{array}{l}
\begin{aligned}
&\frac{{{\partial ^2}L\left( h_2 \right)}}{{{\partial }\tau _l}{{\partial }f _l}} = \\
&\frac{8 \pi^2 nm \Delta f T_s }{{\sigma _{h,l}^2}}\sum\limits_{n,m}^{N,M_s} {{\mathop{\rm Re}\nolimits} \left[ 2 {{\left| {{d_{n,m,l}}} \right|}^2}+ z_{n,m,h}^{l} {d_{n,m,l}} \right]}.
\end{aligned}
\end{array}
\end{equation}

Moreover, the non-diagonal elements of $\boldsymbol{E_{\tau \tau }}$, $\boldsymbol{E_{\tau f }}$ and $\boldsymbol{E_{f f}}$ are all zero.

\ifCLASSOPTIONcaptionsoff
  \newpage
\fi

\end{document}